\documentclass[
 reprint,
 superscriptaddress,
 aps,
 pre,
 longbibliography
]{revtex4-2}

\usepackage{bm}
\usepackage{amsmath}
\usepackage{amsthm}
\usepackage{xcolor} 
\usepackage{array}
\usepackage{amssymb}
\usepackage{graphicx} 
\usepackage{tikz}
\usetikzlibrary{arrows.meta,positioning,calc}
\usepackage[colorlinks=true,allcolors=blue,hypertexnames=false]{hyperref} 

\newcommand{\f}{\frac}

\newcommand{\dd}{\mathop{}\!\mathrm{d}}
\newcommand{\dv}[2]{\frac{\mathrm{d}#1}{\mathrm{d}#2}}
\DeclareMathOperator{\rank}{rank}

\newcommand{\balpha}{\boldsymbol{\alpha}}
\newcommand{\Id}{\mathbb I}
\newcommand{\Z}{\mathbb Z}
\newcommand{\diag}{\operatorname{diag}}

\newcommand{\bms}{\bm s}

\begin{document}

\title{Mutual information in fluctuating hydrodynamics: Role of nonequilibrium long-range correlations}
\author{Rikuya Kobashi}
\email{kobashi.rikuya.84s@st.kyoto-u.ac.jp}
\author{Keiji Saito}
\email{keiji.saitoh@scphys.kyoto-u.ac.jp}
\affiliation{Department of Physics, Kyoto University, Kyoto 606-8502, Japan}
\begin{abstract}
  Nonequilibrium driving can generate correlations that extend throughout a system, even when microscopic interactions are local. How much information do these long-range correlations carry between macroscopic regions? We address this question using linear fluctuating hydrodynamics for a boundary-driven conserved field. We show that, in all physically relevant spatial dimensions, ($d=1,2,3$), nonequilibrium long-range correlations contribute only $O(1)$ to the mutual information between the left and right halves of the system, despite their system-spanning extent. Thus, these correlations generate only a finite amount of shared information as the system size increases. This behavior follows from the inverse-fourth-power decay of the contribution from each nonzero transverse mode; growing mutual information appears only in ($d\ge5$). Our results show that macroscopic spatial correlations can have a limited information-theoretic contribution, consistent with a local-equilibrium description of the bulk.
\end{abstract}

\maketitle

\section{Introduction}
\label{sec:intro}
Information theory connects the statistics of a physical system to its
thermodynamics, even far from equilibrium. In stochastic
thermodynamics~\cite{Sekimoto1998,Seifert2012,VanVu2023}, the Shannon entropy of the system's probability
distribution gives its mean entropy~\cite{Hatano2001,Seifert2005,Esposito2010}. In information
thermodynamics, mutual information links the correlations created by
a measurement to the work that can be extracted through
feedback~\cite{Sagawa2010,Landauer1961,Szilard1929}. The connection extends to quantum thermodynamics,
where resource theories use ideas from quantum information to
characterize which transformations are allowed under thermodynamic
constraints~\cite{Chitambar2019, Brandao2013, Horodecki2013, Shiraishi2025, Tajima2025}. One important direction in nonequilibrium physics
is to identify which information-theoretic quantities matter and to
clarify their thermodynamic meaning.

In quantum information theory, mutual information and entanglement are
used to characterize correlations between two spatial regions, $A$ and
$B$~\cite{Hastings2007, Gioev2006,Bernigau2015,Tajik2023,Eisler2014,GullansHuse2019,Iblisdir2009}. Entanglement measures quantify the nonseparability of the
joint quantum state~\cite{Horodecki2009,Peres1996,Bennett1996,Werner1989,Nielsen1999} while mutual information measures the total
correlations, including both classical and quantum contributions.
For finite-temperature Gibbs states of quantum lattice systems with
finite-range interactions, Wolf et al.\ used the Gibbs variational
principle to show that the mutual information is bounded above by a
constant times the area of the boundary between $A$ and $B$~\cite{Wolf2008}.
This area law is consistent with the behavior of many noncritical
equilibrium states, in which connected correlations between distant
local observables decay exponentially, a property known as exponential
clustering~\cite{Hastings2006,Nachtergaele2006,Kliesch2014,Araki1969, Kim2025}.

Here we study mutual information in classical nonequilibrium steady
states (NESSs), where spatial correlations can differ qualitatively
from those at equilibrium. For example, when the two ends of a system
are coupled to heat reservoirs at different temperatures, the steady
heat current can generate correlations between local energy fluctuations
that extend across the system~\cite{Procaccia1979,Schmitz1985, Kirkpatrick1982, Spohn1983, Garrido1990, Kirkpatrick2013, OrtizDeZarate2006, Saito2021, Bixon1969, Nishikawa2025, Minami2026}. Such long-range correlations can
arise even when the microscopic interactions are local. Similar
correlations occur in diffusive systems in which a chemical potential
difference drives a particle current. These correlations have been
studied using stochastic lattice models, fluctuating hydrodynamics, and
macroscopic fluctuation theory~\cite{Derrida2001,Derrida2002,Santos2001, Derrida2007, Bertini2002,Bertini2005}. In experiments, light scattering
has revealed enhanced long-wavelength temperature fluctuations in a
liquid under a steady temperature gradient~\cite{Lawb1988,Segre1992}.
Microgravity experiments have also observed millimeter-scale
concentration fluctuations in a dilute polymer solution~\cite{Vailati2011}.

How much mutual information do these system-spanning correlations
generate between two macroscopic regions? The information-theoretic
properties of NESSs have been studied in several microscopic models,
including free-fermion systems~\cite{Eisler2014}, diffusive fermion systems
carrying a steady current~\cite{GullansHuse2019}, and the open quantum symmetric simple
exclusion process (quantum SSEP)~\cite{Bernard2023}. In the classical open SSEP,
local equilibrium determines the leading extensive contribution to the
Gibbs--Shannon entropy, even in the presence of long-range correlations.
The correction due to nonequilibrium correlations is
subextensive~\cite{Derrida2007}.
These results concern specific models, and the general size dependence
of mutual information in nonequilibrium transport remains poorly
understood.

As a first step toward unified understanding, we use fluctuating hydrodynamics to describe diffusive
transport of a conserved field. By describing the dynamics of
coarse-grained fields, this framework provides a common description of
a wide range of transport processes and can capture properties that are
insensitive to microscopic details. We ask whether the mutual
information between two macroscopic regions grows in proportion to the
system volume, whether the gapless long-wavelength modes of the
conserved field can give rise to logarithmic growth, or whether the
mutual information can remain bounded.

Specifically, we consider a spatially discretized model of linear fluctuating hydrodynamics for a single conserved field, the local energy
density [Eqs.~(2) and (3)]. In the hydrodynamic framework, each local cell is large compared with the range of microscopic equilibrium correlations and small compared with the length scale over which the mean temperature varies appreciably. Microscopic interactions establish local equilibrium, and their effects enter the hydrodynamic description through transport coefficients and the strength of local equilibrium fluctuations. At this spatial resolution, short-range equilibrium correlations are reflected in the variance of the energy within each cell. The equilibrium covariance in our model is therefore local, so the Gaussian equilibrium distribution factorizes over cells and the
mutual information between disjoint regions vanishes (see Sec.\ref{subsec:equilibrium}). Boundary driving, by contrast, generates correlations between cells that can extend across the system. The mutual information between the two regions thus measures the statistical dependence created by the drive at hydrodynamic scales. We denote it by ${\cal I}_{\mathrm{hyd}}$.

Let $n$ be the number of lattice sites along one side and $h$ the lattice spacing, so that $L=nh$ sets the linear size of the system. As illustrated in Fig.~1, we divide the system into left and right regions by a plane perpendicular to the heat flow. We simply summarize the findings on the mutual information $I_{\mathrm{hyd}}$ as follows. For a fixed nonzero temperature difference between the reservoirs, the mutual information between the two regions scales as
\begin{equation}
{\cal I}_{\mathrm{hyd}}^{(d)}(\mathrm{L}:\mathrm{R}) =
\begin{cases}
\Theta(1), & d<5,\\[3pt]
\Theta\!\left(\log \dfrac{L}{h}\right), & d=5,\\[6pt]
\Theta\!\left[\left(\dfrac{L}{h}\right)^{d-5}\right], & d>5,
\end{cases}
\end{equation}
as $L/h\to\infty$. The mutual information remains bounded for $d<5$, grows logarithmically at $d=5$, and grows as a power law for $d>5$.
In every dimension $d\geq 2$, it grows more slowly than the interfacial area measured in lattice units. In the physically relevant dimensions $d=1,2,3$, the information shared by the two halves thus remains finite as the system grows, even though the correlations extend across the system. In this information-theoretic sense, the result is consistent with a local-equilibrium description of the bulk. This local picture may help deepen our understanding of nonequilibrium thermodynamics.

This paper is organized as follows. In Sec.~II, we introduce the spatially discretized model and derive
the equations and stationary covariances of the transverse modes. We also explain the resulting nonequilibrium long-range correlations and clarify the equilibrium reference state. In Sec.~III, we decompose the mutual information into independent contributions from these modes and show that a single canonical
correlation determines the contribution of each mode. In Sec.~IV, we
present the main results for the individual modes and their sum.
In Sec.~V, we explain the scaling in terms of the range and strength
of the correlations in each mode. In Sec.~VI, we study the two-dimensional boundary-driven SSEP, comparing the exact mutual information with its Gaussian approximation for small systems and examining the system-size dependence of the Gaussian mutual information for larger systems. In Sec.~VII, we prove the scaling
laws by deriving bounds that hold uniformly for all nonzero transverse
modes. In Sec.~VIII, we summarize the results and discuss their
implications.

\begin{figure}
    \hspace*{-0\linewidth}
    \includegraphics[width=0.98\linewidth]{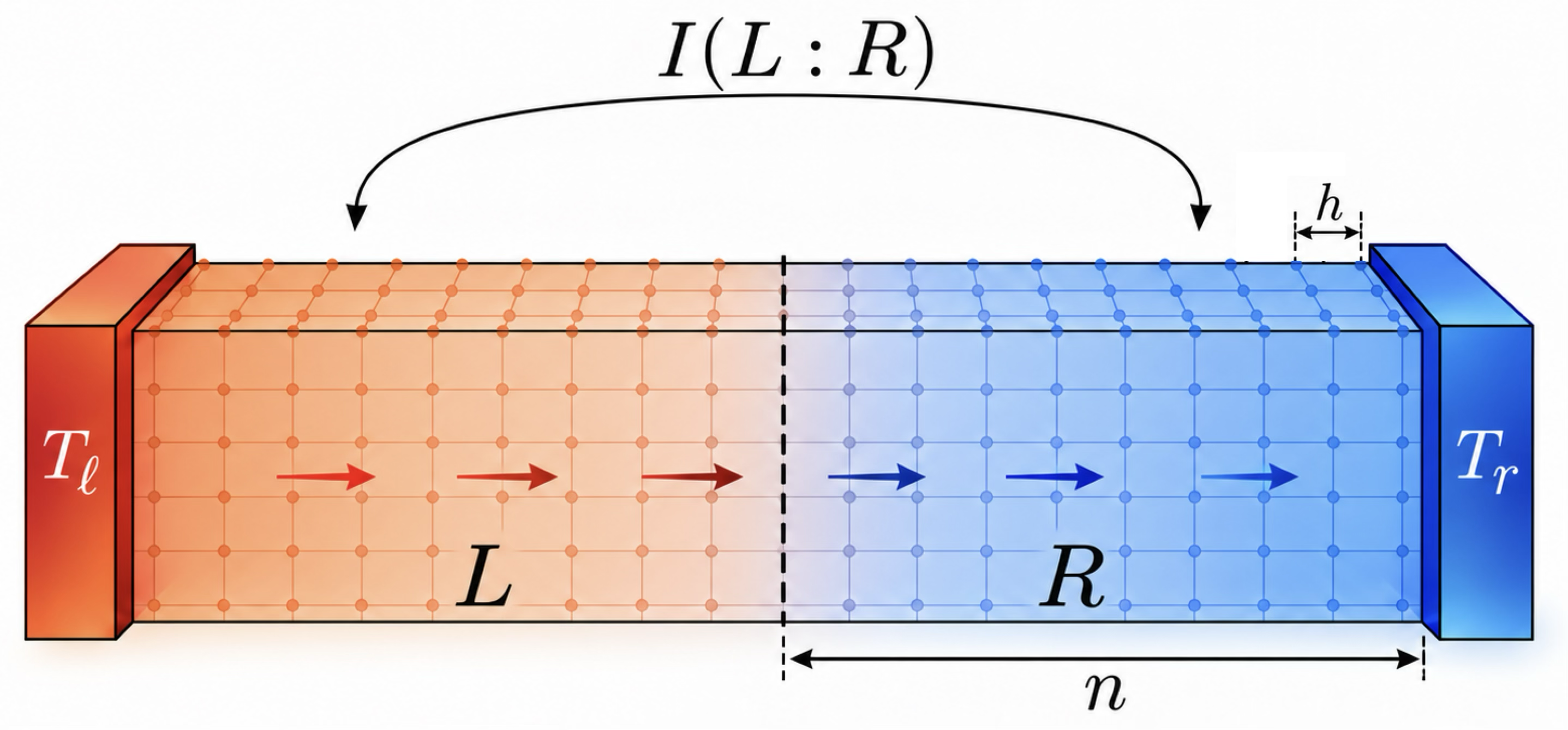}
    \caption{Schematic of the boundary-driven diffusive system and the mutual information $I(L:R)$ between its left and right halves. }
    \label{fig:placeholder}
\end{figure}

\section{Model and transverse-mode equations}
\label{sec:model}

\subsection{Spatially Discretized Fluctuating Hydrodynamics}

We consider a $d$-dimensional lattice with lattice spacing $h$. 
Each cell contains enough particles for local thermodynamics to apply.
Its width $h$ is much larger than the equilibrium correlation length and much
smaller than the length scale over which the mean temperature varies.
We denote the coordinate along the heat-flow direction by $z$ and the transverse coordinate by $\bm y$, with
$z\in h\{1,\ldots,2n+1\},~\bm y \in \{0,h,\ldots,(n-1)h\}^{d-1}$. Each lattice site is labeled by $(z,\bm y)$. For $d=1$, the transverse coordinate $\bm y$ is omitted.

We regard $\hat u_{z,\bm y}(t)$ as the local energy density in a cell of volume $h^d$. Its time evolution obeys the local conservation law
\begin{align}
\frac{\dd\hat u_{z,\bm y}}{\dd t}
={}&
\f{\hat J^{(z)}_{z-h/2,\bm y}
-
\hat J^{(z)}_{z+h/2,\bm y}}{h}
\nonumber\\
&+
\sum_{a=2}^{d}
\f{\left(
\hat J^{(a)}_{z,\bm y-h\bm e_a/2}
-
\hat J^{(a)}_{z,\bm y+h\bm e_a/2}
\right)}{h}.
\label{eq:localConservation}
\end{align}
Here, $\bm e_a$ denotes the unit vector
along the $a$th transverse direction
($a=2,\ldots,d$). The energy currents in the heat-flow direction and in the $a$-th transverse direction are defined, respectively, as
\begin{align}
\hat J^{(z)}_{z+h/2,\bm y}(t)
&:=
\f{D}{h}\bigl(
\hat u_{z,\bm y}-\hat u_{z+h,\bm y}
\bigr)
\nonumber\\
&\quad
+\sqrt{\f{2\kappa}{h^d}}\,
\theta_{z+h/2}
\hat\xi^{(z)}_{z+h/2,\bm y}(t),
\nonumber\\
\hat J^{(a)}_{z,\bm y+h\bm e_a/2}(t)
&:=
\f{D}{h}\bigl(
\hat u_{z,\bm y}
-
\hat u_{z,\bm y+h\bm e_a}
\bigr)
\nonumber\\
&\quad
+\sqrt{\f{2\kappa}{h^d}}\,
T_z
\hat\xi^{(a)}_{z,\bm y+h\bm e_a/2}(t),
\qquad a=2,\ldots,d,
\label{eq:energyCurrents}
\end{align}
where $D>0$ is the diffusion coefficient and $\kappa>0$ characterizes the strength of thermal fluctuations. The first term represents the diffusive current, while the second represents the stochastic current arising from thermal fluctuations. We place heat baths at $z=0$ and $z=N+1$ with temperatures $T_\ell$ and $T_r$, respectively, and set the specific heat to unity. The boundary conditions are then given by $
        \hat u_{0,\bm y}(t)
        =
        T_\ell,
        \hat u_{N+1,\bm y}(t)
        =
        T_r,
        \hat u_{z,\bm y}(t)
        =
        \hat u_{z,\bm y+hn\bm e_a}(t), ~
        (a=2,\ldots,d)$.
Under these boundary conditions, the steady-state temperature profile is
\begin{align}
T_z
&=
T_\ell+z\delta_n,
\quad
\delta_n:=
\frac{T_r-T_\ell}{N+1},
\label{eq:Tprofile}.
\end{align}
The noise amplitude on a bond in the heat-flow direction is chosen as
\begin{align}
\theta_{z+1/2}^{2}
:=
\frac{T_z^2+T_{z+1}^2}{2},
\qquad
z=0,1,\ldots,N.
\label{eq:theta}
\end{align}
We denote the standard Gaussian white noise on each bond by $\hat\xi_\lambda(t)$, with the composite index $\lambda=(\mu,z,\bm y)$, and define
$\hat\xi_{1,z,\bm y}:=\hat\xi^{(z)}_{z+1/2,\bm y}$ and
$\hat\xi_{a,z,\bm y}:=\hat\xi^{(a)}_{z,\bm y+\bm e_a/2}$ for $a=2,\ldots,d$.
These noises satisfy
\begin{align}
\left\langle \hat\xi_\lambda(t)\right\rangle
&=
0,
\qquad
\left\langle
\hat\xi_\lambda(t)
\hat\xi_{\lambda'}(t')
\right\rangle
=
\delta_{\lambda\lambda'}\delta(t-t'),
\end{align}
where
\begin{align}
\delta_{\lambda\lambda'}
&:=
\delta_{\mu\mu'}\delta_{zz'}\delta_{\bm y\bm y'},
\qquad
\delta_{\bm y\bm y'}
:=
\prod_{a=2}^{d}\delta_{y_a y'_a}.
\end{align}

For notational simplicity, we introduce the rescaling
\begin{equation}
    \begin{split}
        &h^{d/2}\delta\hat u_{z,\bm y}(t)
        \mapsto
        \delta\hat u_{z/h,\bm y/h}(h^{-2}t),\\
        &h\hat\xi_\lambda(t)
        \mapsto
        \hat\xi_{\lambda/h}(h^{-2}t),\\
        &(h^{-2}t,z/h,\bm y/h)
        \mapsto
        (t,z,\bm y).
    \end{split}
    \label{eq:scaletrans}
\end{equation}
Here, the fluctuation around the steady-state mean is defined  $
\delta\hat u_{z,\bm y}(t)
:=
\hat u_{z,\bm y}(t)
-
\langle\hat u_{z,\bm y}(t)\rangle_{\mathrm{ns}}.$
$X\mapsto Y$ denotes the replacement of $X$ by $Y$ in
Eqs.~\eqref{eq:localConservation} and \eqref{eq:energyCurrents}.
Under these replacements, Eqs.~\eqref{eq:localConservation} and
\eqref{eq:energyCurrents} can be written as
\begin{align}
\frac{\dd \delta \hat u_{z,\bm y}}{\dd t}
={}&
D\bigl(
\delta  \hat u_{z-1,\bm y}
-
2\delta \hat u_{z,\bm y}
+
\delta  \hat u_{z+1,\bm y}
\bigr)
\nonumber\\
&+
D\sum_{a=2}^{d}
\bigl(
\delta \hat u_{z,\bm y-\bm e_a}
-
2\delta \hat u_{z,\bm y}
+
\delta \hat u_{z,\bm y+\bm e_a}
\bigr)
\nonumber\\
&+
\sqrt{2\kappa}\,
\hat\xi_{z,\bm y}(t),
\label{eq:localDynamics}
\end{align}
where $\hat\xi_{z,\bm y}(t)$ denotes the site noise at $(z,\bm y)$,
defined as the difference between the incoming and outgoing bond noises:
\begin{align}
\hat\xi_{z,\bm y}(t)
:={}&
\theta_{z-1/2}
\hat\xi^{(z)}_{z-1/2,\bm y}(t)
-
\theta_{z+1/2}
\hat\xi^{(z)}_{z+1/2,\bm y}(t)
\nonumber\\
&+
T_z
\sum_{a=2}^{d}
\left[
\hat\xi^{(a)}_{z,\bm y-\bm e_a/2}(t)
-
\hat\xi^{(a)}_{z,\bm y+\bm e_a/2}(t)
\right].
\label{eq:siteNoise}
\end{align}

For later use, we rewrite Eq.~\eqref{eq:localDynamics} in vector form.
The fluctuations around the steady-state mean satisfy the homogeneous Dirichlet boundary conditions $
\delta\hat u_{0,\bm y}=\delta\hat u_{N+1,\bm y}=0$ along the heat-flow direction.
Accordingly, we define the positive discrete Laplacian
$L_N^{\mathrm D}\in\mathbb R^{N\times N}$,
which incorporates the homogeneous Dirichlet boundary conditions and is restricted to the interior sites
$z,z'=1,\ldots,N$, by
\begin{equation}
(L_N^{\mathrm D})_{zz'}
=
\begin{cases}
2, & z=z',\\
-1, & |z-z'|=1,\\
0, & \text{otherwise}.
\end{cases}
\label{eq:longitudinalLaplacian}
\end{equation}
For $n\ge3$, we also define the positive periodic discrete Laplacian
$\Delta_{\perp,n}\in\mathbb R^{n\times n}$ in one transverse direction as
\begin{equation}
(\Delta_{\perp,n})_{yy'}
=
\begin{cases}
2, & y=y',\\[1mm]
-1, & |y-y'|=1,\\[1mm]
-1, & (y,y')=(0,n-1)\ \text{or}\ (n-1,0),\\[1mm]
0, & \text{otherwise}.
\end{cases}
\label{eq:transverseLaplacian}
\end{equation}

In what follows, operators acting on multiple transverse directions are represented using the Kronecker product $\otimes$. The periodic discrete Laplacian acting on the $(d-1)$ transverse directions is therefore
\begin{align}
\Delta_{\perp,n}^{(d-1)}
:=
\sum_{a=2}^{d}
\Id_n^{\otimes(a-2)}
\otimes
\Delta_{\perp,n}
\otimes
\Id_n^{\otimes(d-a)}.
\label{eq:transverseLaplacianD}
\end{align}
Here, $\Id_m$ denotes the $m\times m$ identity matrix, and
$\Id_n^{\otimes m}$ denotes the tensor product of $m$ copies of $\Id_n$.
The positive discrete Laplacian on the full lattice is then
\begin{equation}
\Delta_{d,n}
:=
L_N^{\mathrm D}
\otimes
\Id_n^{\otimes(d-1)}
+
\Id_N
\otimes
\Delta_{\perp,n}^{(d-1)}.
\label{eq:fullLaplacian}
\end{equation}

We define the fluctuation vector on the full lattice by
\begin{align}
\delta\hat{\bm u}(t)
:=
\bigl(
\delta\hat{u}_{z,\bm y}(t)
\bigr)_{(z,\bm y)}
\in
\mathbb R^{Nn^{d-1}},
\label{eq:fluctuationVector}
\end{align}
and similarly arrange the site noises defined in Eq.~\eqref{eq:siteNoise} in the same order as
\begin{align}
\hat{\bm\xi}(t)
:=
\bigl(
\hat\xi_{z,\bm y}(t)
\bigr)_{(z,\bm y)}
\in
\mathbb R^{Nn^{d-1}}.
\label{eq:siteNoiseVector}
\end{align}
Equation~\eqref{eq:localDynamics} can then be written as
\begin{align}
\frac{\dd\,\delta\hat{\bm u}(t)}{\dd t}
=
-D\Delta_{d,n}\,
\delta\hat{\bm u}(t)
+
\sqrt{2\kappa}\,
\hat{\bm\xi}(t).
\label{eq:fluctuationDynamicsVector}
\end{align}
Equation~\eqref{eq:fluctuationDynamicsVector} separates the spatial dependence in the heat-flow and transverse directions into
$L_N^{\mathrm D}$ and $\Delta_{\perp,n}^{(d-1)}$, respectively.
We exploit this structure in the next section by decomposing the transverse dependence into Fourier modes.

\subsection{Transverse-Mode Decomposition}
\label{subsec:fourier}
We now Fourier transform Eq.~\eqref{eq:fluctuationDynamicsVector} in the transverse directions. 
For a single transverse direction, we define the index set of real modes $\mathcal A_n$ as
\begin{equation}
\mathcal A_n
:=
\{0\}
\cup
\{(m,c),(m,s):m=1,\ldots,M\}
\label{eq:AnOdd}
\end{equation}
for odd $n=2M+1$, and as
\begin{equation}
\mathcal A_n
:=
\{0,\pi\}
\cup
\{(m,c),(m,s):m=1,\ldots,M-1\}
\label{eq:AnEven}
\end{equation}
for even $n=2M$.
In both cases, $|\mathcal A_n|=n$.
For each $\alpha\in\mathcal A_n$, we define a real vector
$w_\alpha\in\mathbb R^n$ by
\begin{align}
w_\alpha(y):=
\begin{cases}
n^{-1/2}, & \alpha=0,\\
\sqrt{2/n}\cos(2\pi my/n), & \alpha=(m,c),\\
\sqrt{2/n}\sin(2\pi my/n), & \alpha=(m,s),\\
(-1)^y n^{-1/2}, & \alpha=\pi \quad (n\ \mathrm{even}).
\end{cases}
\label{eq:realModes}
\end{align}
These vectors form a real orthonormal basis and satisfy
$
\Delta_{\perp,n}w_\alpha
=
\rho_\alpha w_\alpha.$
The corresponding eigenvalues are
\begin{align}
\rho_\alpha:=
\begin{cases}
0, & \alpha=0,\\
4\sin^2(\pi m/n), & \alpha=(m,c),(m,s),\\
4, & \alpha=\pi \quad (n\ \mathrm{even}).
\end{cases}
\label{eq:eigenvalues}
\end{align}
For $d\ge2$, we label the real transverse modes by
$
\boldsymbol\alpha
=
(\alpha_2,\ldots,\alpha_d)
\in
\mathcal A_n^{d-1}.$ The corresponding eigenvector and eigenvalue are defined by
\begin{align}
w_{\boldsymbol\alpha}(\bm y)
:=
\prod_{a=2}^{d}
w_{\alpha_a}(y_a),
\qquad
\rho_{\boldsymbol\alpha}
:=
\sum_{a=2}^{d}
\rho_{\alpha_a},
\label{eq:multiMode}
\end{align}
so that $
\Delta_{\perp,n}^{(d-1)}
w_{\boldsymbol\alpha}
=
\rho_{\boldsymbol\alpha}
w_{\boldsymbol\alpha}.$
The total number of real transverse modes is
$|\mathcal A_n^{d-1}|=n^{d-1}$.

By Fourier transforming Eq.~\eqref{eq:fluctuationDynamicsVector} using the real basis in Eq.~\eqref{eq:realModes}, the $d$-dimensional fluctuating energy-field dynamics decomposes into independent equations for each real transverse mode:
\begin{equation}
\dv{\delta\hat{\widetilde{\bm u}}_{\boldsymbol\alpha}}
{t}
=
-D\Delta_{\boldsymbol\alpha}
\delta\hat{\widetilde{\bm u}}_{\boldsymbol\alpha}
+
\sqrt{2\kappa}\,
\hat{\bm\eta}_{\boldsymbol\alpha}(t).
\label{eq:modeDynamics}
\end{equation}
Here, we define the Fourier component of mode $\boldsymbol\alpha$ and the corresponding vector collecting its components along the heat-flow direction as
\begin{align}
\begin{split}
\delta\hat{\widetilde u}_{\boldsymbol\alpha,z}
&:=
\sum_{\bm y\in\Z_n^{d-1}}
w_{\boldsymbol\alpha}(\bm y)
\delta\hat u_{z,\bm y},\\
\delta\hat{\widetilde{\bm u}}_{\boldsymbol\alpha}
&:=
\bigl(
\delta\hat{\widetilde u}_{\boldsymbol\alpha,1},
\ldots,
\delta\hat{\widetilde u}_{\boldsymbol\alpha,N}
\bigr)^{\mathsf T}
\in\mathbb R^N.
\end{split}
\label{eq:modeVariables}
\end{align}
Similarly, the mode-resolved white noise along the heat-flow direction is defined by
\begin{align}
\begin{split} 
\hat{\eta}_{\boldsymbol\alpha,z}(t)
&:=
\sum_{\bm y\in\Z_n^{d-1}}
w_{\boldsymbol\alpha}(\bm y)
\hat\xi_{z,\bm y}(t),\\
\hat{\bm\eta}_{\boldsymbol\alpha}(t)
&:=
\bigl(
\hat{\eta}_{\boldsymbol\alpha,1}(t),
\ldots,
\hat{\eta}_{\boldsymbol\alpha,N}(t)
\bigr)^{\mathsf T},
\end{split}
\label{eq:modeNoiseDef}
\end{align}
and the mode-dependent operator is
\begin{equation}
\Delta_{\boldsymbol\alpha}
:=
L_N^{\mathrm D}
+
\rho_{\boldsymbol\alpha}\Id_N.
\label{eq:DeltaAlpha}
\end{equation}

We define the diagonal matrix $F:=\diag(T_1^2,\ldots,T_N^2) \in\mathbb R^{N\times N}$. By performing a straightforward calculation on Eq. \eqref{eq:DeltaAlpha}, the two-point correlation of $\hat{\bm\eta}_{\boldsymbol\alpha}(t)$ is
\begin{equation}
\left\langle
\hat{\bm\eta}_{\boldsymbol\alpha}(t)
\hat{\bm\eta}_{\boldsymbol\beta}^{\mathsf T}(t')
\right\rangle
=
\frac{\delta_{\boldsymbol\alpha\boldsymbol\beta}}{2}
\left(
\Delta_{\boldsymbol\alpha}F
+
F\Delta_{\boldsymbol\alpha}
+
2\delta_n^2\Id_N
\right)
\delta(t-t').
\label{eq:modeNoiseCovariance}
\end{equation}
Here,$
\delta_{\boldsymbol\alpha\boldsymbol\beta}
:=
\prod_{a=2}^{d}
\delta_{\alpha_a\beta_a}.$ In particular, the Gaussian noises driving different transverse modes are independent.

Since $\Delta_{\boldsymbol\alpha}>0$, Eq.~\eqref{eq:modeDynamics} admits a unique zero-mean Gaussian stationary state. We define the stationary covariance matrix of mode $\boldsymbol\alpha$ by
\begin{equation}
(C_{\boldsymbol\alpha})_{ij}
:=
\left\langle
\delta\hat{\widetilde{\bm u}}_{\boldsymbol\alpha, i}
\delta\hat{\widetilde{\bm u}}_{\boldsymbol\alpha,j}^{\mathsf T}
\right\rangle_{\mathrm{ns}}.
\label{eq:SigmaAlpha}
\end{equation}
From the linear Langevin equation~\eqref{eq:modeDynamics} and the noise covariance~\eqref{eq:modeNoiseCovariance}, the stationary covariance satisfies the Lyapunov equation
\begin{equation}
D\Delta_{\boldsymbol\alpha}C_{\boldsymbol\alpha}
+
DC_{\boldsymbol\alpha}\Delta_{\boldsymbol\alpha}
=
\kappa
\left(
\Delta_{\boldsymbol\alpha}F
+
F\Delta_{\boldsymbol\alpha}
+
2\delta_n^2\Id_N
\right).
\label{eq:Lyapunov}
\end{equation}
Its unique solution is
\begin{equation}
C_{\boldsymbol\alpha}
=
\frac{\kappa}{D}
\left(
F
+
\delta_n^2\Delta_{\boldsymbol\alpha}^{-1}
\right).
\label{eq:modeCovariance}
\end{equation}
From Eq.~\eqref{eq:modeNoiseCovariance} together with the block-diagonal drift matrix, the unique stationary Gaussian states associated with different real transverse modes are mutually independent. A verification that the stationary cross-covariances vanish is given in Appendix~\ref{app:independence}.

\subsection{Nonequilibrium long-range correlation}
\label{subsec:nonequilibriumCorrelations}

Equation~\eqref{eq:modeCovariance} separates the stationary fluctuations into
a local-equilibrium contribution and a correction generated
by the temperature difference between the reservoirs.
The diagonal term $(\kappa/D)F$ gives the local-equilibrium
contribution to the variance at the temperature of each
cell. This term is local in real space and therefore does
not correlate different cells. The remaining term,
$(\kappa/D)\delta_n^2\Delta_{\boldsymbol\alpha}^{-1}$, modifies the local
variances and introduces correlations between different
cells. Since $\delta_n=(T_r-T_\ell)/(N+1)$, this correction
is proportional to the square of the temperature difference
at fixed system size and vanishes when the two reservoirs
have the same temperature.

The spatial structure of this correction is determined
by $\Delta_{\boldsymbol\alpha}^{-1}$, the Green function of the diffusion
operator for mode $\boldsymbol\alpha$. Its form reveals how local
diffusive dynamics can produce correlations over
macroscopic distances. This is particularly clear in one
dimension, where $\Delta_{\boldsymbol 0}=L_N^{\mathrm D}$.
Using the rescaled variables introduced in Eq.~\eqref{eq:scaletrans},
consider two distinct sites $i<j$ and define their macroscopic
positions by $x=i/(N+1)$ and $x'=j/(N+1)$. The Green function
of the Dirichlet Laplacian gives
\begin{equation}
  \left\langle
    \delta\hat u_i\,\delta\hat u_j
  \right\rangle_{\mathrm{ns}}
  =
  \frac{\kappa}{D}
  \frac{(T_r-T_\ell)^2}{N+1}
  x(1-x').
  \label{eq:nonequilibriumCorrelation1D}
\end{equation}

The factor $x(1-x')$ describes a correlation profile that
extends across the entire sample. If $x$ and $x'$ are held
fixed in the interior as the system grows, the distance
between the two cells increases in proportion to $N$.
Their covariance retains the same spatial profile, with
an amplitude that decreases as $N^{-1}$ at fixed boundary
temperatures. The spatial extent and the strength of the
correlations thus have different size dependences:
the former grows with the system, while the latter
decreases for each pair of cells at distinct macroscopic
positions. These weak correlations over macroscopic
distances are the nonequilibrium long-range correlations
considered here.

The same Green-function structure appears in higher
dimensions after the transverse-mode decomposition.
The zero transverse mode retains the one-dimensional
Dirichlet Green function. For a nonzero transverse mode,
the additional term $\rho_{\boldsymbol\alpha}\Id_N$ in
$\Delta_{\boldsymbol\alpha}$ limits the range of the correlations along
the heat-flow direction. For fixed transverse wave-vector
indices, however, $\rho_{\boldsymbol\alpha}=O(n^{-2})$, and the
corresponding longitudinal range remains of order $n$.
Thus, the lowest nonzero transverse modes also carry
correlations whose spatial extent grows with the system size.

The presence of correlations across the entire sample
raises a further question: how much information do they
carry between two macroscopic regions? Their spatial
extent alone does not answer this question, because their
strength and the local fluctuations also enter the
statistical dependence between the regions. In
Sec.~\ref{sec:channel}, we use the full stationary covariance
to calculate the mutual information between the left and
right halves of the system.
Before turning to this calculation, we clarify the equilibrium
properties of the present model.

\subsection{Equilibrium case}
\label{subsec:equilibrium}

Fluctuating hydrodynamics describes fields obtained by
averaging over cells that contain many microscopic degrees
of freedom. Each cell is large enough for local thermodynamics
to apply, yet small compared with the macroscopic length
scales of interest. We therefore assume
$\xi_{\mathrm{eq}}\ll h\ll \ell_{\mathrm{hyd}}$, where
$\xi_{\mathrm{eq}}$ is the microscopic equilibrium correlation
length, $h$ is the cell size, and $\ell_{\mathrm{hyd}}$ is a
macroscopic length scale. Local equilibration within each
cell is also assumed to be fast compared with the evolution
of the hydrodynamic fields. The cells can exchange energy
through their boundaries while remaining locally equilibrated.

At this spatial resolution, microscopic equilibrium
correlations enter primarily through local thermodynamic
quantities and the fluctuation strength of each cell.
Coarse-graining alone does not make neighboring cells
exactly independent, because microscopic degrees of freedom
near a cell boundary can remain correlated across it.
The present model retains the local contribution to the
equilibrium covariance and neglects these residual
correlations between cells. Its equilibrium properties
should be understood within this local approximation.

Consider equal reservoir temperatures,
$T_\ell=T_r=T>0$. The stationary temperature profile is then
uniform, and the nonequilibrium contribution in
Eq.~\eqref{eq:modeCovariance} vanishes. The covariance of
every transverse mode becomes
$C_{\boldsymbol\alpha}=\sigma_T^2\Id_N$, where
$\sigma_T^2:=\kappa T^2/D$. Since the transverse modes are
independent and have the same covariance, the inverse
orthogonal transformation gives a real-space covariance
$\sigma_T^2\Id_{M_{\mathrm{cell}}}$, with
$M_{\mathrm{cell}}=Nn^{d-1}$ the total number of cells.
Thus, each cell retains thermal fluctuations, while the
equal-time covariance between different cells vanishes.

The corresponding Fokker--Planck equation makes the
structure of the equilibrium distribution explicit.
Let $\boldsymbol X=(X_1,\ldots,X_{M_{\mathrm{cell}}})$ denote
the real-space energy-density fluctuations after the rescaling in
Eq.~\eqref{eq:scaletrans}, and let $P(\boldsymbol X,t)$ be
their probability density. At equilibrium,
\begin{equation}
  \partial_t P
  =
  D\sum_{a,b=1}^{M_{\mathrm{cell}}}
  (\Delta_{d,n})_{ab}\,
  \partial_{X_a}
  \left(
    X_b P+\sigma_T^2\partial_{X_b}P
  \right).
  \label{eq:equilibriumFokkerPlanck}
\end{equation}
Its stationary solution is the product Gaussian distribution
\begin{equation}
  P_{\mathrm{eq}}(\boldsymbol X)
  =
  \prod_{a=1}^{M_{\mathrm{cell}}}
  \frac{1}{\sqrt{2\pi\sigma_T^2}}
  \exp\!\left(
    -\frac{X_a^2}{2\sigma_T^2}
  \right).
  \label{eq:equilibriumProduct}
\end{equation}
Indeed,
$X_bP_{\mathrm{eq}}
+\sigma_T^2\partial_{X_b}P_{\mathrm{eq}}=0$
for every $b$, so the probability current vanishes.
The reservoir boundary conditions imply
$\Delta_{d,n}>0$, which ensures that this stationary
distribution is unique. Although diffusion and the
stochastic currents couple neighboring cells, their
fluctuation-dissipation balance produces an equilibrium
distribution that factorizes over cells.

Equation~\eqref{eq:equilibriumProduct} also factorizes
between the left and right regions. Consequently,
\begin{equation}
  \mathcal I_{\mathrm{hyd}}^{\mathrm{eq}}
  (\mathrm L:\mathrm R)=0.
  \label{eq:equilibriumMI}
\end{equation}
This is an exact statement about the equal-time equilibrium
distribution of the present model. It does not imply that
the cells evolve independently: fluctuations can propagate
through diffusion, and correlations between different
cells at different times can be nonzero.

The vanishing of
$\mathcal I_{\mathrm{hyd}}^{\mathrm{eq}}$
is consistent with the coarse-grained description adopted
here. Microscopic degrees of freedom on opposite sides
of the interface can remain correlated at equilibrium,
and these short-distance correlations can contribute
to the mutual information of the microscopic system.
Such contributions are not retained in the local
approximation to the equilibrium covariance used in
this model. A nonzero microscopic equilibrium mutual
information is therefore compatible with the product
distribution in Eq.~\eqref{eq:equilibriumProduct}.

The equilibrium state considered here is maintained by
reservoirs at a common temperature, so the total energy
of the system is allowed to fluctuate. Imposing an exact
constraint on the total energy would couple the cell
variables and alter the factorization above. Likewise,
the assumed separation $\xi_{\mathrm{eq}}\ll h$ need not
hold near an equilibrium critical point, where spatial
correlations must be retained explicitly. A formal
continuum limit $h\to0$ of the discretized equations is
a limit of the effective theory; its physical use
presumes that microscopic and hydrodynamic scales remain
well separated. Within these assumptions, the equilibrium
product state provides the reference for the correlations
and mutual information generated by nonequilibrium driving.

\section{Mode-resolved mutual information and single-channel structure}
\label{sec:channel}
\subsection{Mode Decomposition of the Mutual Information}

We first divide the system at the central cross section and define the left and right regions as $
L_n
:=
\{1,\ldots,n\}\times\Z_n^{d-1},
R_n :=
\{n+1,\ldots,N\}\times\Z_n^{d-1}.$ We denote the local energy variables contained in the left and right regions by
\begin{align}
\hat{\bm u}_{L} :=
\bigl(
\hat u_{z,\bm y}
\bigr)_{(z,\bm y)\in L_n}, \quad 
\hat{\bm u}_{R} :=
\bigl(
\hat u_{z,\bm y}
\bigr)_{(z,\bm y)\in R_n},
\label{eq:LRenergyVectors}
\end{align}
respectively. We denote the mutual information between $L_n$ and $R_n$ in the stationary state by $\mathcal{I}_{\mathrm{hyd}}^{(d)}(n)$. Let
$p_{\mathrm{ns}}(\bm u_L,\bm u_R)$ be the stationary joint probability density of
$\hat{\bm u}_{L}$ and $\hat{\bm u}_{R}$, and let
$p_{\mathrm{ns}}^L(\bm u_L)$ and
$p_{\mathrm{ns}}^R(\bm u_R)$ be the corresponding marginal probability densities.
Then $\mathcal{I}_{\mathrm{hyd}}^{(d)}(n)$ is defined as
\begin{align}
\mathcal{I}_{\mathrm{hyd}}^{(d)}(n)
:=
\int
\dd\bm u_L\,\dd\bm u_R\,
p_{\mathrm{ns}}(\bm u_L,\bm u_R)
\log
\frac{
p_{\mathrm{ns}}(\bm u_L,\bm u_R)
}{
p_{\mathrm{ns}}^L(\bm u_L)
p_{\mathrm{ns}}^R(\bm u_R)
}.
\label{eq:totalMIDef}
\end{align}

Next, we consider the transverse Fourier transformation introduced in Sec.~\ref{sec:model}. This transformation acts separately within the left $L_n$ and right $R_n$ regions as an invertible orthogonal transformation and therefore does not mix the degrees of freedom across the bipartition. Since mutual information is invariant under invertible transformations applied independently to the two subsystems, the real-space mutual information $\mathcal{I}_{\mathrm{hyd}}^{(d)}(n)$ is equal to that computed from the transverse Fourier variables.

For each mode $\boldsymbol\alpha$, we define the variables in the left half as
\begin{align}
\hat{\bm X}_{\boldsymbol\alpha}
:=
\bigl(
\delta\hat{\widetilde u}_{\boldsymbol\alpha,1},
\ldots,
\delta\hat{\widetilde u}_{\boldsymbol\alpha,n}
\bigr)^{\mathsf T}
\in\mathbb R^n,
\label{eq:Xalpha}
\end{align}
and those in the right half as
\begin{align}
\hat{\bm Y}_{\boldsymbol\alpha}
:=
\bigl(
\delta\hat{\widetilde u}_{\boldsymbol\alpha,n+1},
\delta\hat{\widetilde u}_{\boldsymbol\alpha,n+2},
\ldots,
\delta\hat{\widetilde u}_{\boldsymbol\alpha,N}
\bigr)^{\mathsf T}
\in\mathbb R^n.
\label{eq:Yalpha}
\end{align}
The mutual information carried by mode $\boldsymbol\alpha$ is then defined as
\begin{align}
\mathcal{I}_{\mathrm{hyd},\boldsymbol\alpha}^{(d)}(n)
:=
I(
\hat{\bm X}_{\boldsymbol\alpha}
:
\hat{\bm Y}_{\boldsymbol\alpha}
).
\label{eq:Ialpha}
\end{align}

As shown in Sec.~\ref{sec:model}, the stationary distributions of distinct real transverse modes are mutually independent. Therefore, using the invariance of mutual information under the transverse Fourier transformation together with its additivity over independent modes, we obtain
\begin{align}
\mathcal{I}_{\mathrm{hyd}}^{(d)}(n)
=
\sum_{\boldsymbol\alpha\in\mathcal A_n^{d-1}}
\mathcal{I}_{\mathrm{hyd},\boldsymbol\alpha}^{(d)}(n).
\label{eq:MIsum}
\end{align}
In the next subsection, we evaluate $\mathcal{I}_{\mathrm{hyd},\boldsymbol\alpha}^{(d)}(n)$.

\subsection{Rank-One Cross-Covariance and Single-Channel Structure}
As obtained in Eq.~\eqref{eq:modeCovariance}, the stationary distribution of each mode is Gaussian. Hence, the joint distribution of
$(\hat{\bm X}_{\boldsymbol\alpha},\hat{\bm Y}_{\boldsymbol\alpha})$
is also Gaussian, and its mutual information is fully determined by the covariance matrix. For subsequent analysis, let the covariance matrix be expressed as follows:
\begin{align}
C_{\boldsymbol\alpha}
=\frac{\kappa\delta_n^2}{D}\left(
Q+\Delta_{\boldsymbol\alpha}^{-1} \right).
\label{eq:GammaAlpha}
\end{align}
Here,
\begin{align}
Q
&:=
\diag\left(
\left(\frac{T_1}{\delta_n}\right)^2,\ldots,
\left(\frac{T_N}{\delta_n}\right)^2
\right).
\label{eq:Qdef}
\end{align}
We partition $C_{\boldsymbol\alpha}$ as
\begin{align}
C_{\boldsymbol\alpha}
=
\frac{\kappa\delta_n^2}{D}
\begin{pmatrix}
C^{(LL)}_{\boldsymbol\alpha}
&
C^{(LR)}_{\boldsymbol\alpha}
\\
(C^{(LR)}_{\boldsymbol\alpha})^{\mathsf T}
&
C^{(RR)}_{\boldsymbol\alpha}
\end{pmatrix},
\label{eq:LRblock}
\end{align}
where
$C^{(LL)}_{\boldsymbol\alpha},C^{(RR)}_{\boldsymbol\alpha},C^{(LR)}_{\boldsymbol\alpha}\in\mathbb R^{n\times n}$
denote the normalized covariance blocks within the left region, within the right region, and between the two regions, respectively.

Since the differential entropy of a Gaussian distribution is determined solely by the determinant of its covariance matrix, the definition
$I(X:Y)=h(X)+h(Y)-h(X,Y)$
gives
\begin{align}
\mathcal{I}_{\mathrm{hyd},\boldsymbol\alpha}^{(d)}(n)
=
\frac12
\log
\frac{
\det C^{(LL)}_{\boldsymbol\alpha}
\det C^{(RR)}_{\boldsymbol\alpha}
}{
\det \left[\frac{D}{\kappa\delta_n^2}C_{\boldsymbol\alpha} \right]
}.
\label{eq:GaussianMI}
\end{align}

The results obtained so far follow generally from the Gaussian nature of the model. We now exploit the specific structure of the present model to make the information-theoretic content of Eq.~\eqref{eq:GaussianMI} explicit. We first derive an explicit expression for $\Delta_{\boldsymbol\alpha}^{-1}$, which determines the correlation structure between the left and right regions. For $\rho_{\boldsymbol\alpha}\ge0$, we define $\gamma_{\boldsymbol\alpha}\ge0$ by
\begin{equation}
\cosh\gamma_{\boldsymbol\alpha}
=
1+\frac{\rho_{\boldsymbol\alpha}}{2},
\label{eq:gamma}
\end{equation}
and introduce the sequence
\begin{equation}
s_{\boldsymbol\alpha,k}
:=
\begin{cases}
\dfrac{
\sinh(k\gamma_{\boldsymbol\alpha})
}{
\sinh\gamma_{\boldsymbol\alpha}
},
&
\gamma_{\boldsymbol\alpha}>0,
\\[2mm]
k,
&
\gamma_{\boldsymbol\alpha}=0,
\end{cases}
\qquad
k=0,1,\ldots,N+1.
\label{eq:salpha}
\end{equation}
In terms of $s_{\boldsymbol\alpha,k}$, we have
\begin{equation}
(\Delta_{\boldsymbol\alpha}^{-1})_{ij}
=
\frac{
s_{\boldsymbol\alpha,\min(i,j)}
s_{\boldsymbol\alpha,N+1-\max(i,j)}
}{
s_{\boldsymbol\alpha,N+1}
},
\label{eq:GreenKernel}
\end{equation}
as can be directly verified by $\Delta_{\boldsymbol\alpha}^{-1}\Delta_{\boldsymbol\alpha} = \Id_N$. For the zero transverse mode $\boldsymbol\alpha=\boldsymbol 0$, we have
$\rho_{\boldsymbol 0}=0$ and
$s_{\boldsymbol 0,k}=k$, so that
\begin{equation}
(\Delta_{\boldsymbol 0}^{-1})_{ij}
=
\frac{
\min(i,j)
[N+1-\max(i,j)]
}{
N+1
}.
\label{eq:tentKernel}
\end{equation}
For fixed $j$, this kernel increases linearly from the left boundary up to $i=j$ and then decreases linearly toward the right boundary, forming a tent-shaped profile. Thus, the correlations associated with the zero transverse mode extend over the entire system along the heat-flow direction.

We now consider the left-right cross-covariance block
$C^{(LR)}_{\boldsymbol\alpha}$ defined in Eq.~\eqref{eq:LRblock}.
Since $Q$ is diagonal, it does not contribute to the cross-covariance block.
For $i,k=1,\ldots,n$, we have
\begin{align}
(C^{(LR)}_{\boldsymbol\alpha})_{ik}
&=
(\Delta_{\boldsymbol\alpha}^{-1})_{i,n+k}
\nonumber\\
&=
\frac{
s_{\boldsymbol\alpha,i}
s_{\boldsymbol\alpha,n+1-k}
}{
s_{\boldsymbol\alpha,N+1}
},
\label{eq:Ccomponent}
\end{align}
where we used $N=2n$ and $i\le n<n+k$.
Introducing the column vectors
\begin{align}
\bms_{\boldsymbol\alpha}
&:=
\bigl(
s_{\boldsymbol\alpha,1},
\ldots,
s_{\boldsymbol\alpha,n}
\bigr)^{\mathsf T},
\qquad
\bm t_{\boldsymbol\alpha}
:=
\bigl(
s_{\boldsymbol\alpha,n},
\ldots,
s_{\boldsymbol\alpha,1}
\bigr)^{\mathsf T},
\label{eq:sVector}
\end{align}
we obtain
\begin{equation}
C^{(LR)}_{\boldsymbol\alpha}
=
\frac{
\bms_{\boldsymbol\alpha}
\bm t_{\boldsymbol\alpha}^{\mathsf T}
}{
s_{\boldsymbol\alpha,N+1}
},
\qquad
\rank C^{(LR)}_{\boldsymbol\alpha}
=
1.
\label{eq:rankOne}
\end{equation}
Thus, the cross-covariance block $C_{\balpha}$ connecting the left and right regions has rank one.
This means that, for each transverse mode, there is only one correlation channel connecting the two regions.
Using the Schur complement and the matrix determinant lemma in
Eqs.~\eqref{eq:GaussianMI} and \eqref{eq:rankOne}, we obtain
\begin{align}
\mathcal{I}_{\mathrm{hyd},\boldsymbol\alpha}^{(d)}(n)
&=
-\frac12
\log
\left(
1-\sigma_{\boldsymbol\alpha}^2
\right),
\nonumber\\
\sigma_{\boldsymbol\alpha}^2
&=
\frac{
\bms_{\boldsymbol\alpha}^{\mathsf T}
(C^{(LL)}_{\boldsymbol\alpha})^{-1}
\bms_{\boldsymbol\alpha}
}{
s_{\boldsymbol\alpha,N+1}
}
\frac{
\bm t_{\boldsymbol\alpha}^{\mathsf T}
(C^{(RR)}_{\boldsymbol\alpha})^{-1}
\bm t_{\boldsymbol\alpha}
}{
s_{\boldsymbol\alpha,N+1}
}.
\label{eq:canonicalCorrelation}
\end{align}
Equivalently, $\sigma_{\boldsymbol\alpha}\in[0,1)$ is the unique nonzero singular value of
\begin{equation}
(C^{(LL)}_{\boldsymbol\alpha})^{-1/2}
C^{(LR)}_{\boldsymbol\alpha}
(C^{(RR)}_{\boldsymbol\alpha})^{-1/2}.
\label{eq:singularvalue}
\end{equation}
The statistical dependence between the left and right regions for each real transverse mode is completely characterized by a single information channel, as a direct consequence of the rank-one structure of $C^{(LR)}_{\boldsymbol\alpha}$ in Eq.~\eqref{eq:rankOne}.

Finally, we note that the mutual informations defined in
Eqs.~\eqref{eq:totalMIDef} and \eqref{eq:Ialpha}
are invariant under the rescaling in Eq.~\eqref{eq:scaletrans}.
This follows because the Jacobian factors appearing in the probability densities and integration measures cancel in the definition~\eqref{eq:totalMIDef}.

\section{Main results}
\label{sec:mainResults}

In this section, we present our main results. In the following, we take the system size to be $2n$ in the heat-flow direction and $n$ in each transverse direction, and consider a fixed spatial dimension $d$ and fixed boundary temperatures $0<T_r <T_\ell$. For a real transverse mode $\boldsymbol\alpha$, we denote the corresponding integer wave-vector index by $\bm m=(m_2,\ldots,m_d)$ and define $r=|\bm m|$.

The first main result is that the mutual information carried by a nonzero transverse mode decays as $r^{-4}$ with the transverse wave number. Specifically, there exist positive constants $c_I$ and $C_I$, independent of $n$ and $\boldsymbol\alpha$, such that for every nonzero transverse mode,
\begin{align}
\frac{c_I}{r^4}
\le
\mathcal{I}_{\mathrm{hyd},\boldsymbol\alpha}^{(d)}(n)
\le
\frac{C_I}{r^4}.
\label{eq:mainModeMI}
\end{align}
Here, $c_I$ and $C_I$ are proportional to $(T_r-T_\ell)^4$. In contrast, the zero transverse mode contributes
$\mathcal{I}_{\mathrm{hyd},\bm 0}^{(d)}(n)=\Theta(1)$.

The second main result is that, by combining the decay of $\mathcal{I}_{\mathrm{hyd},\boldsymbol\alpha}^{(d)}(n)$ with the growth in the number of transverse modes, the total mutual information between the left and right regions exhibits the dimension-dependent scaling
\begin{align}
\mathcal{I}_{\mathrm{hyd}}^{(d)}(n)=
\begin{cases}
\Theta(1), & d\le4,\\
\Theta(\log n), & d=5,\\
\Theta(n^{d-5}), & d\ge6.
\end{cases}
\label{eq:mainTotalMI}
\end{align}

The first physical implication of our result is that the presence of long-range correlations does not necessarily imply a large amount of information shared between two macroscopic regions. Indeed, for $d\ge2$, the mutual information obtained here satisfies
\begin{align}
\mathcal{I}_{\mathrm{hyd}}^{(d)}(n)=o(n^{d-1}),
\end{align}
and is therefore asymptotically smaller than the area of the interface separating the two regions. Although there are $n^{d-1}$ independent transverse modes, the contribution of each nonzero mode is sufficiently suppressed because it decays as $r^{-4}$, as shown in Eq.~\eqref{eq:mainModeMI}. The distinction between the spatial extent of correlations and the amount of mutual information they carry becomes even clearer in one dimension. In this case, the left-right correlations are carried solely by the zero mode, whose correlations extend throughout the system along the heat-flow direction. Nevertheless,
\begin{align}
\mathcal{I}_{\mathrm{hyd}}^{(1)}(n)=\Theta(1),
\end{align}
so that the mutual information remains of the same order as a one-dimensional area law. This behavior is also consistent with known results for one-dimensional boundary-driven diffusive systems such as the open SSEP \cite{Derrida2007}. In the present model, the mutual information measures the nonadditive part of the differential entropy between the left and right regions, and the above result therefore implies that this nonadditivity vanishes at leading order.

Next, the dependence of $\mathcal{I}_{\mathrm{hyd}}^{(d)}(n)$ on microscopic degrees of freedom can be made explicit by restoring the lattice spacing $h$. Let the system size in each transverse direction be $L:=hn$. Then, from Eq.~\eqref{eq:mainTotalMI}, the scaling of the mutual information in the continuum limit $h\to0$ at fixed $L$, can be written as
\begin{align}
\mathcal{I}_{\mathrm{hyd}}^{(d)}(n)
=
\begin{cases}
\Theta(1), & d<5,\\
\Theta\!\left(\log \dfrac{L}{h}\right), & d=5,\\
\Theta\!\left[\left(\dfrac{L}{h}\right)^{d-5}\right], & d>5.
\end{cases}
\end{align}
Here, $f=\Theta(g)$ means that $f$ is bounded above and
below by positive constant multiples of $g$ in the stated limit. We have used the invariance of mutual information under the rescaling in Eq.~\eqref{eq:scaletrans}.

This representation reveals that the degrees of freedom governing the mutual information. For $d<5$, the leading behavior of the mutual information remains finite as $h\to0$. High-wave-number modes therefore do not affect the leading scaling, which can be described entirely in terms of long-wavelength hydrodynamic degrees of freedom. By contrast, at $d=5$,
\begin{align}
\mathcal{I}_{\mathrm{hyd}}^{(5)}(n)\sim A_5\log\frac{L}{h},
\end{align}
showing a logarithmic dependence on the microscopic cutoff. Here, $f\sim g$ denotes asymptotic equivalence,
i.e., $f/g\to1$ in the stated limit. This is the marginal case in which long- and short-wavelength scales contribute on an equal logarithmic footing. In particular, the coefficient $A_5$ can be determined from the physical argument developed in Sec.~\ref{sec:Physicalinterp}. Thus, for $d<5$, the mutual information, and at $d=5$, the coefficient $A_5$, are universal quantities independent of microscopic details. For $d>5$, on the other hand,
\begin{align}
\mathcal{I}_{\mathrm{hyd}}^{(d)}(n) = 
\Theta\left[\left(\frac{L}{h}\right)^{d-5} \right],
\end{align}
so that the leading behavior depends algebraically on the cutoff, reflecting a dominant contribution from high-wave-number degrees of freedom. 

Finally, near equilibrium, the mutual information first appears at fourth order in the temperature gradient. Indeed, the left-right covariance generated by the nonequilibrium driving is second order in the temperature gradient, and the corresponding $\sigma_{\balpha}$ scales as
\begin{align}
\sigma_{\boldsymbol\alpha}
=
O\!\left(
\left(\frac{\delta_n}{h}\right)^2
\right).
\end{align}
For $\sigma_{\boldsymbol\alpha}\ll1$, $
\mathcal{I}_{\mathrm{hyd}, \boldsymbol\alpha}^{(d)}(n)
\simeq \sigma_{\boldsymbol\alpha}^2/2$, hence
\begin{align}
\mathcal{I}_{\mathrm{hyd}, \boldsymbol\alpha}^{(d)}(n)
=
O\!\left(
\left(\frac{\delta_n}{h}\right)^4
\right).
\end{align}
Thus, the mutual information appears at quadratic order in its amplitude under weak nonequilibrium driving.

\section{Physical picture of mutual-information scaling}
\label{sec:Physicalinterp}

To elucidate the physical mechanism governing the scaling of
$\mathcal{I}_{\mathrm{hyd}}^{(d)}(n)$, we first show, prior to the rigorous analysis, how the same scaling follows from an intuitive argument. The behavior of the mutual information carried by each transverse mode can be understood in terms of the correlation length along the heat-flow direction and the strength of the corresponding correlations. 

We first consider the correlation length. For mode $\boldsymbol\alpha$, the correlation length is given by
$\gamma_{\boldsymbol\alpha}^{-1}$, which characterizes the distance over which correlations induced by the transverse mode propagate along the heat-flow direction. Increasing the transverse wave number therefore acts as an effective mass in the heat-flow direction and localizes the left-right correlations near the interface. This behavior can be seen directly from the Green function in Eq.~\eqref{eq:GreenKernel}. We consider a mode $\boldsymbol\alpha$ satisfying $r\gg1$. For two points located $\ell$ sites to the left and $q$ sites to the right of the interface, namely $i=n-\ell$ and $j=n+1+q$, the cross-covariance takes the form
\begin{align}
    \begin{split}
        \left(\Delta_{\boldsymbol\alpha}^{-1}\right)_{n-\ell,\,n+1+q}
        &\sim
        a_{\balpha}u_{\ell}u_q,\\
        a_{\boldsymbol\alpha}
        &=
        \frac{e^{-\gamma_{\boldsymbol\alpha}}}
        {2\sinh\gamma_{\boldsymbol\alpha}},
    \end{split}
\label{eq:GreenPhysicalForm}
\end{align}
in the regime where the effects of the outer boundaries can be neglected. Here, $u_{\ell}=e^{-\gamma_{\balpha}\ell}.$

It follows that the degrees of freedom relevant to the mutual information are localized within a distance of order the correlation length
$\gamma_{\balpha}^{-1}$ from the interface. Since $\sigma_{\balpha}$ is the unique nonzero singular value of the matrix in Eq.~\eqref{eq:singularvalue}, we obtain
\begin{align}
    \sigma_{\balpha}
    =
    \left\|
    (C^{(LL)}_{\boldsymbol\alpha})^{-1/2}
    C^{(LR)}_{\boldsymbol\alpha}
    (C^{(RR)}_{\boldsymbol\alpha})^{-1/2}
    \right\|
    \sim
    \f{a_{\balpha}\delta_n^2}{T_c^2}
    \|u\|^2.
\label{eq:sigmaPhyscal}
\end{align}
Here, $\|\cdot\|$ denotes the operator norm when acting on a matrix and the $L^2$ norm when acting on a vector. In Eq.\eqref{eq:sigmaPhyscal}, the off-diagonal elements of
$A_{\balpha}$ and $B_{\balpha}$ can be neglected, and the correlations relevant to the mutual information are localized near the interface. We therefore replace the local temperatures near the cut by
$T_c:=(T_r+T_{\ell})/2$. For matrix elements whose two indices
are both near the cut, we estimate
\begin{align}
C^{(LL)}_{\boldsymbol\alpha}
\sim
\f{T_c^2}{\delta_n^2} I,
\qquad
C^{(RR)}_{\boldsymbol\alpha}
\sim
\f{T_c^2}{\delta_n^2} I.
\end{align}
Interpreting $\|u\|^2$ as the effective number of degrees of freedom $n^{\mathrm{eff}}$, involved in the correlation, Eq.~\eqref{eq:sigmaPhyscal} can be written as
\begin{align}
    \sigma_{\balpha}
    \sim
    \f{
    a_{\balpha}\delta_n^2
    (n^{\mathrm{eff}})^2
    }{
    T_c^2 n^{\mathrm{eff}}
    }.
\end{align}
This expression admits a simple interpretation: the numerator is the product of the correlation strength and the effective number of correlated pairs, whereas the denominator represents the self-correlation strength multiplied by the number of participating degrees of freedom. 

The effective number of degrees of freedom is related to the correlation length
$\xi_{\boldsymbol\alpha}:=\gamma_{\boldsymbol\alpha}^{-1}$ through
\begin{align}
    n^{\mathrm{eff}}
    =
    \|u\|^2
    =
    \sum_{\ell=0}^{n-1}
    e^{-2\gamma_{\balpha}\ell}
    \sim
    \f{\xi_{\boldsymbol\alpha}}{2}.
\end{align}
We therefore obtain
\begin{align}
    \sigma_{\boldsymbol\alpha}
    \sim
    \frac{
    a_{\boldsymbol\alpha}\delta_n^2
    }{
    2T_c^2
    }
    \xi_{\boldsymbol\alpha}.
\label{eq:correlationPhysical}
\end{align}

We first consider the low-wave-number modes with $r=O(1)>0$. In this regime,
\begin{align}
\gamma_{\boldsymbol\alpha}
=
O(n^{-1}),
\qquad
\xi_{\boldsymbol\alpha}
=
O(n),
\end{align}
and the typical correlation amplitude is of order $O(n)$. Accordingly,
\begin{align}
a_{\boldsymbol\alpha}
=
O(n).
\end{align}
It then follows that
\begin{align}
\sigma_{\boldsymbol\alpha}
=
O(1),
\qquad
\mathcal{I}_{\mathrm{hyd}, \boldsymbol\alpha}^{(d)}(n)
=
O(1).
\label{eq:lowModeMIPhysical}
\end{align}
Thus, low-wave-number modes carry a finite amount of mutual information.

In the intermediate long-wavelength regime $1\ll r\ll n$, we have
\begin{align}
\gamma_{\boldsymbol\alpha}
\simeq
\frac{2\pi r}{n},
\qquad
\xi_{\boldsymbol\alpha}
\sim
\frac{n}{2\pi r}.
\label{eq:intermediateCorrelationLength}
\end{align}
In this regime, the amplitude of $C_{\balpha}$ near the interface is of order $O(n/r)$, and hence
\begin{align}
a_{\boldsymbol\alpha}
\sim
\f{1}{2\gamma_{\balpha}}
\sim
\f{n}{4\pi r}.
\end{align}
Thus, as the transverse wave number increases, not only does the correlation between a given pair of sites become weaker, but the spatial region over which the correlations extend also shrinks as $n/r$. Combining these two effects gives
\begin{align}
\sigma_{\boldsymbol\alpha}
\sim
\frac{(T_r-T_{\ell})^2}{64\pi^2T_c^2r^2}
=: \frac{A_{\sigma}}{r^2}.
\label{eq:intermediateSigmaPhysical}
\end{align}
Furthermore, since $\sigma_{\boldsymbol\alpha}\ll1$,
\begin{align}
\begin{split}
\mathcal{I}_{\mathrm{hyd}, \boldsymbol\alpha}^{(d)}(n)
&\simeq
\frac12
\sigma_{\boldsymbol\alpha}^{2}\\
&\sim
\frac{A^2_{\sigma}}{2r^4}.
\end{split}
\label{eq:intermediateMIPhysical}
\end{align}

Finally, for the high-wave-number modes with $r=\Theta(n)$, both
$\rho_{\boldsymbol\alpha}$ and $\gamma_{\boldsymbol\alpha}$ are generically of order unity, so that
\begin{align}
\xi_{\boldsymbol\alpha}
=
O(1).
\end{align}
Accordingly,
\begin{align}
a_{\boldsymbol\alpha}
=
O(1).
\end{align}
It then follows that
\begin{align}
\sigma_{\boldsymbol\alpha}
=
O(n^{-2}),
\qquad
\mathcal{I}_{\mathrm{hyd}, \boldsymbol\alpha}^{(d)}(n)
=
O(n^{-4}).
\label{eq:highModeMIPhysical}
\end{align}
Thus, high-wave-number modes behave as very weak information channels that are strongly localized near the interface.

The scaling behaviors obtained above for the different wave-number regimes are summarized in Table~\ref{tab:modeScalingSummary}. Summing these mode contributions yields the dimension-dependent scaling of the total mutual information $\mathcal{I}_{\mathrm{hyd}}^{(d)}(n)$. The same argument also shows that the scaling is unchanged whether the boundary temperatures $T_r$ and $T_\ell$ are kept fixed or scaled as $O(n)$ so that the temperature gradient $\delta_n$ remains fixed.

\begin{table}[t]
\caption{Scaling of $\mathcal{I}_{\mathrm{hyd}, \boldsymbol\alpha}^{(d)}(n)$ in different $\balpha$ regimes}
\label{tab:mode_scaling}
\begin{ruledtabular}
\begin{tabular}{ccccc}
$\balpha$ regime
& $\xi_{\balpha}$
& $a_{\balpha}$
& $\sigma_{\balpha}$
& $\mathcal{I}_{\mathrm{hyd}, \boldsymbol\alpha}^{(d)}(n)$
\\
\hline
$r=O(1)$
& $O(n)$
& $O(n)$
& $O(1)$
& $O(1)$
\\
$1\ll r\ll n$
& $O(n/r)$
& $O(n/r)$
& $O(r^{-2})$
& $O(r^{-4})$
\\
$r=\Theta(n)$
& $O(1)$
& $O(1)$
& $O(n^{-2})$
& $O(n^{-4})$
\end{tabular}
\end{ruledtabular}
\label{tab:modeScalingSummary}
\end{table}

Summing the mode contributions in Eq.~\eqref{eq:MIsum}, we estimate
\begin{align}
\mathcal{I}_{\mathrm{hyd}}^{(d)}(n)
&\sim
\frac{A_{\sigma}^2}{2}
\sum_{\bm m\neq\bm 0}
|\bm m|^{-4}
+
I_{\bm 0}(L:R)
\nonumber\\
&\sim
\frac{A_{\sigma}^2}{2}
S_{d-2}
\int_1^n
\dd r\,
r^{d-6}
+
O(1),
\label{eq:modeIntegralPhysical}
\end{align}
which reproduces the scaling of the total mutual information in
Eq.~\eqref{eq:mainTotalMI}. Here,
\begin{align}
S_{d-2}
:=
\frac{2\pi^{(d-1)/2}}
{\Gamma((d-1)/2)}
\end{align}
denotes the surface area of the $(d-2)$-dimensional unit sphere.

\begin{figure}[t]
    \centering
    \hspace*{-0.08\linewidth}
    \includegraphics[width=1.05\linewidth, trim=0 40 0 0]{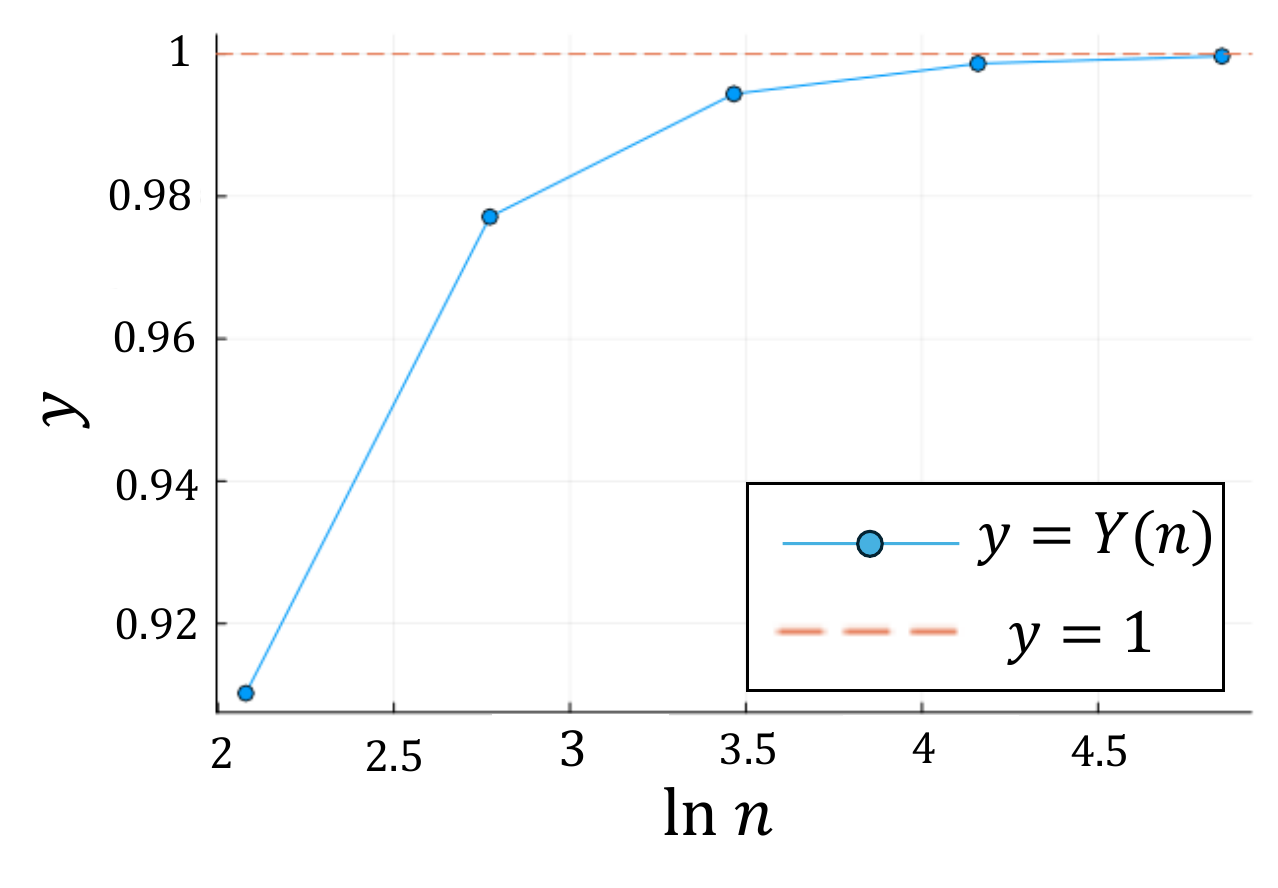}
    \caption{
    Numerical verification of the logarithmic coefficient at $d=5$. In the numerical calculations, the system size $n$ was increased from $8$ to $256$, doubling at each step.
    }
    \label{fig:A5_numerics}
\end{figure}

As discussed in Sec.~\ref{sec:mainResults}, for $d=5$ the logarithmic coefficient is
\begin{align}
A_5
=
\frac{A_{\sigma}^2}{2}S_3
=
\f{1}{4096\pi^2}
\left(
\frac{T_r-T_{\ell}}{T_c}
\right)^4.
\label{eq:A5value}
\end{align}
We verify this prediction numerically in Fig.~\ref{fig:A5_numerics}. For each transverse mode, we numerically evaluate $\mathcal{I}_{\mathrm{hyd}, \boldsymbol\alpha}^{(d)}(n)$ using Eq.~\eqref{eq:canonicalCorrelation}, and then obtain the total mutual information $\mathcal{I}_{\mathrm{hyd}}^{(d)}(n)$ from the mode sum in Eq.~\eqref{eq:MIsum}. We define
\begin{align}
Y(n)
:=
\frac{J(2n)-J(n)}{\log 2},
\quad
J(n)
:=
\frac{
\mathcal{I}_{\mathrm{hyd}}^{(d)}(n)-\mathcal{I}_{\mathrm{hyd}, \bm 0}^{(d)}(n)
}{
A_5\left(2n/(2n+1)\right)^4
}.
\end{align}
If the logarithmic coefficient of the mutual information is indeed given by $A_5$ in Eq.~\eqref{eq:A5value}, then $Y(n)$ approaches unity as $n$ increases. As shown in Fig.~\ref{fig:A5_numerics}, the numerical results converge toward $1$ with increasing system size, confirming the analytical prediction
$I_n=A_5\log n+O(1)$ and, in particular, the value of the logarithmic coefficient $A_5$.

The above argument combines the asymptotic behavior of $\mathcal{I}_{\mathrm{hyd}, \boldsymbol\alpha}^{(d)}(n)$ in each wave-number regime with the density number of modes. In the following sections, we establish the scaling rigorously by deriving uniform upper and lower bounds on $\sigma_{\balpha}$ of each mode for finite $n$.

\section{Comparison with the Two-Dimensional SSEP}

In this work, we have shown that the mutual information between the left and right halves of a two-dimensional system remains \(O(1)\) in the system size within linear fluctuating hydrodynamics. The same scaling behavior is expected to hold in microscopic models governed by similar diffusive hydrodynamics. To examine this expectation, we consider the two-dimensional symmetric simple exclusion process (SSEP), a representative microscopic model of diffusive transport. Indeed, Landim et al. showed that the stationary density fluctuation field of the boundary-driven SSEP converges to a Gaussian field in the hydrodynamic limit \cite{Landim2008}, and also discussed extensions to higher dimensions. Motivated by this correspondence, we numerically compare the Shannon mutual information computed directly from the stationary distribution of the SSEP with the Gaussian mutual information obtained from its stationary two-point covariance matrix. We further examine the system-size dependence of the Gaussian mutual information to test its consistency with the two-dimensional scaling predicted by our fluctuating-hydrodynamic theory.

We consider the two-dimensional SSEP on the same lattice as in Sec. II, \(\Lambda_n=\{1,\ldots,2n\}\times\mathbb Z_n\). Each site \(i=(z,y)\) is characterized by an occupation variable \(\eta_i\in\{0,1\}\). We divide the system into the left and right regions, \(L_n=\{1,\ldots,n\}\times\mathbb Z_n\) and \(R_n=\{n+1,\ldots,2n\}\times\mathbb Z_n\), respectively. In the bulk, for each nearest-neighbor pair \((i,j)\) with different occupations, the two occupations are exchanged during an infinitesimal time interval \(dt\) with probability \(dt+o(dt)\). We impose periodic boundary conditions in the transverse direction and couple the two ends along the particle-flow direction to particle reservoirs with densities \(\rho_\ell\) and \(\rho_r\), respectively. The Markov generator of this process is given by

\[
\begin{aligned}
(\mathcal L f)(\eta)
={}&\sum_{\langle i,j\rangle}
\bigl[f(\eta^{i,j})-f(\eta)\bigr]\\
&+\sum_{s=\ell,r}\sum_{y\in\mathbb Z_n}
c_s(\eta_{z_s,y})
\bigl[f(\eta^{z_s,y})-f(\eta)\bigr].
\end{aligned}
\]
Here, $\eta^{i,j}$ denotes the configuration obtained from $\eta$ by exchanging the occupations at sites $i$ and $j$, while $\eta^i$ denotes the configuration obtained from $\eta$ by flipping the occupation at site $i$. We set \(z_\ell=1\), \(z_r=2n\), and \(c_s(a)=\rho_s(1-a)+(1-\rho_s)a\). The stationary mean density profile is given by
\[
\bar\rho_z=\rho_\ell+\frac{z(\rho_r-\rho_\ell)}{2n+1}.
\]

In the numerical calculations, we set \(\rho_\ell=0.2\) and \(\rho_r=0.8\). For small systems (\(n\leq 3\)), we directly solve the stationary equation for the Markov generator to obtain the stationary distribution, from which we compute the mutual information \(\mathcal{I}_{\mathrm{SSEP}}^{(d)}(n)\) between the left and right regions. For larger systems, we evaluate the Gaussian mutual information from the stationary two-point covariance matrix \(C_{\mathrm{SSEP}}\) as

\[
\mathcal{I}_{\mathrm{G}}^{(d)}(n)
=
\frac12\log
\frac{\det C_{\mathrm{SSEP}}^{LL}\det C_{\mathrm{SSEP}}^{RR}}
{\det C_{\mathrm{SSEP}}}.
\]

By comparing these quantities, we examine the accuracy of the Gaussian approximation at finite system sizes and the system-size dependence of the Gaussian mutual information.

\begin{figure}
    \centering
    \hspace*{-0.08\linewidth}
    \includegraphics[width=1\linewidth]{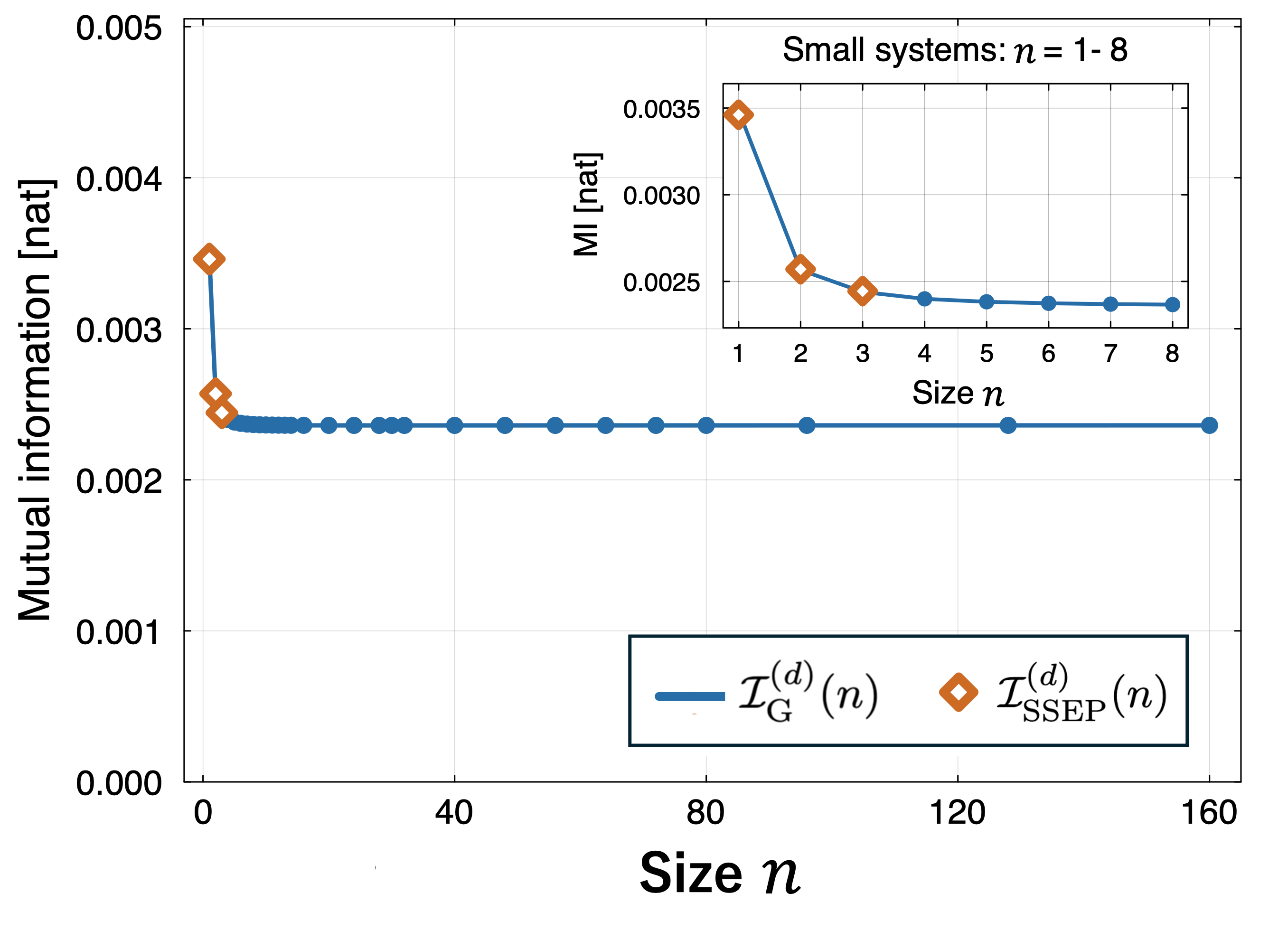}
    \caption{System-size dependence of the mutual information between the left and right halves of the two-dimensional SSEP on a $2n\times n$ lattice with reservoir densities $\rho_\ell=0.2$ and $\rho_r=0.8$. The solid line represents the Gaussian mutual information calculated from the stationary two-point covariance matrix, while the symbols represent the Shannon mutual information computed directly from the stationary distribution for $n\leq 3$. The inset shows an enlargement of the small-system-size region.}
    \label{fig:placeholder}
\end{figure}

Figure \ref{fig:placeholder} shows the system-size dependence of the mutual information between the left and right halves of the two-dimensional SSEP. For small systems (\(n\leq 3\)), the Shannon mutual information computed directly from the stationary distribution agrees well with the Gaussian mutual information obtained from the two-point covariance matrix. For larger system sizes, the Gaussian mutual information approaches a finite, nonzero value as \(n\) increases. This behavior is consistent with the scaling \(\mathcal{I}_{\mathrm{hyd}}^{(d)}(n)=\Theta(1)\) predicted by linear fluctuating hydrodynamics in two dimensions. Furthermore, the good agreement between the Shannon and Gaussian mutual information at small system sizes suggests that the same scaling behavior may also hold for the exact mutual information of the microscopic SSEP.

\section{Rigorous proof of the mutual-information scaling}
\label{sec:rigorousProof}

To justify the physical picture developed in the previous section, we show that the $r^{-4}$ decay of the mode-resolved mutual information $\mathcal{I}_{\mathrm{hyd}, \boldsymbol\alpha}^{(d)}(n)$ holds uniformly for all nonzero transverse modes at finite $n$. Specifically, it suffices to prove that there exist positive constants $c_{\sigma}$ and $C_{\sigma}$, independent of $n$ and $\balpha$, such that
\begin{align}
\frac{c_{\sigma}}{r^2}
\le
\sigma_{\balpha}
\le
\frac{C_{\sigma}}{r^2},
\qquad
r=|\bm m|.
\end{align}

Using the representation obtained in Sec.~III, the wave-number dependence of $\sigma_{\balpha}$ can be reduced to the estimation of a single scalar quantity. First, the left and right covariance blocks can be written as
\begin{align}
A_{\balpha}
=
Q_L+G_{\balpha}^{LL},
\qquad
B_{\balpha}
=
Q_R+G_{\balpha}^{RR},
\qquad
G_{\balpha}:=\Delta_{\balpha}^{-1}.
\end{align}
Since $G_{\balpha}$ is positive definite and
$\Delta_{\balpha}\ge L_N^{\mathrm D}$, we have the uniform bound
\begin{align}
\|G_{\balpha}\|
\le
\|(L_N^{\mathrm D})^{-1}\|
=
O(n^2)
\end{align}
for all transverse modes. On the other hand, since
$T_\ell\le T_z\le T_r$ and $\delta_n=O(n^{-1})$,
the diagonal matrices $Q_L$ and $Q_R$ are also of order $O(n^2)$.
Combining these observations, we obtain the uniform bounds
\begin{align}
\frac{T_\ell^2}{\delta_n^2}I
\le
A_{\balpha},B_{\balpha}
\le
\frac{M_T}{\delta_n^2}I,
\qquad
M_T
:=
T_r^2+\frac{(T_r-T_\ell)^2}{4},
\end{align}
where the constants are independent of $n$ and $\balpha$.
Therefore,
\begin{align}
\frac{\delta_n^2}{M_T}I
\le
A_{\balpha}^{-1},B_{\balpha}^{-1}
\le
\frac{\delta_n^2}{T_\ell^2}I,
\end{align}
and hence
\begin{align}
\bm s_{\balpha}^{\mathsf T}
A_{\balpha}^{-1}
\bm s_{\balpha}
&\asymp
\delta_n^2
\|\bm s_{\balpha}\|^2,
\nonumber\\
\bm t_{\balpha}^{\mathsf T}
B_{\balpha}^{-1}
\bm t_{\balpha}
&\asymp
\delta_n^2
\|\bm t_{\balpha}\|^2.
\end{align}
Here, $f\asymp g$ has the same meaning as $f=\Theta(g)$,
with bounds uniform in $n$ and $\boldsymbol\alpha$.
Using $\|\bm s_{\balpha}\|=\|\bm t_{\balpha}\|$ together with the expression derived in Sec.~III,
\begin{align}
\sigma_{\balpha}^2
=
\frac{
\bm s_{\balpha}^{\mathsf T}
A_{\balpha}^{-1}
\bm s_{\balpha}
}{
s_{\balpha,2n+1}
}
\frac{
\bm t_{\balpha}^{\mathsf T}
B_{\balpha}^{-1}
\bm t_{\balpha}
}{
s_{\balpha,2n+1}
},
\end{align}
we obtain
\begin{align}
\sigma_{\balpha}
\asymp
\delta_n^2
R_n(\gamma_{\balpha}),
\qquad
R_n(\gamma)
:=
\frac{
\sum_{k=1}^{n}
s_k(\gamma)^2
}{
s_{2n+1}(\gamma)
}.
\end{align}
Here, $s_k(\gamma)$ is the sequence defined in Eq.~\eqref{eq:salpha}.
Thus, determining the wave-number dependence of $\sigma_{\balpha}$ reduces to estimating $R_n(\gamma_{\balpha})$.

The quantity $R_n(\gamma_{\balpha})$ can be evaluated analytically, yielding
\begin{align}
R_n(\gamma)
=
\frac{1}{4\sinh^2\gamma}
\left[
1-
\frac{(2n+1)\sinh\gamma}
{\sinh((2n+1)\gamma)}
\right].
\end{align}
For every nonzero transverse mode, $(2n+1)\gamma_{\balpha}$ is bounded from below by a positive constant independent of $n$ and $\balpha$. Hence, the term in square brackets admits a uniform positive lower bound $b_0>0$, and therefore
\begin{align}
\frac{b_0}{4\sinh^2\gamma_{\balpha}}
\le
R_n(\gamma_{\balpha})
\le
\frac{1}{4\sinh^2\gamma_{\balpha}}
\end{align}
for all nonzero transverse modes.

On the other hand,
\begin{align}
\sinh^2\gamma_{\balpha}
=
\rho_{\balpha}
\left(
1+\frac{\rho_{\balpha}}{4}
\right),
\end{align}
and the transverse eigenvalue satisfies the uniform estimate
\begin{align}
\rho_{\balpha}
\asymp
\frac{r^2}{n^2},
\qquad
r=|\bm m|.
\end{align}
It follows that
\begin{align}
R_n(\gamma_{\balpha})
\asymp
\frac{n^2}{r^2}.
\end{align}
Since $\delta_n^2\asymp n^{-2}$ for fixed boundary temperatures, combining this result with the estimate in the previous paragraph gives
\begin{align}
\sigma_{\balpha}
\asymp
\frac{1}{r^2}.
\end{align}

We now return to the mode-resolved mutual information. Since $\sigma_{\balpha}$ is uniformly bounded away from $1$ for all nonzero transverse modes,
\begin{align}
\mathcal{I}_{\mathrm{hyd}, \boldsymbol\alpha}^{(d)}(n)
=
-\frac{1}{2}\log\!\left(1-\sigma_{\balpha}^2\right)
\asymp
\sigma_{\balpha}^2
\end{align}
holds uniformly. Therefore, using $\sigma_{\balpha}\asymp r^{-2}$ obtained above, we find
\begin{align}
\mathcal{I}_{\mathrm{hyd}, \boldsymbol\alpha}^{(d)}(n)
\asymp
\frac{1}{r^4}.
\end{align}
For the zero transverse mode $\balpha=0$, on the other hand,
\begin{align}
\sigma_0=\Theta(1),
\qquad
I_0(L:R)=\Theta(1).
\end{align}
Thus, the mode-resolved bound in Eq.~\eqref{eq:mainModeMI} is established for all nonzero transverse modes. Substituting this result into the mode-additivity relation
\begin{align}
\mathcal{I}_{\mathrm{hyd}}^{(d)}(n)
=
\sum_{\balpha}
\mathcal{I}_{\mathrm{hyd}, \boldsymbol\alpha}^{(d)}(n)
\end{align}
and counting the number of modes at each wave number then yields the dimension-dependent scaling in Eq.~\eqref{eq:mainTotalMI}. The detailed derivations of the uniform bounds and of the zero-mode contribution are given in Appendix~\ref{app:modeMIBound}.

\section{Summary and discussion}
\label{sec:summary}

In this work, we investigated the mutual information generated between two macroscopic regions by long-range correlations in nonequilibrium steady states, using linear fluctuating hydrodynamics for diffusive transport. We obtained
\begin{align}
\mathcal{I}_{\mathrm{hyd}}^{(d)}(n)
=
\begin{cases}
\Theta(1), & d<5,\\
A_5\log n+O(1), & d=5,\\
\Theta(n^{d-5}), & d>5.
\end{cases}
\end{align}
We also confirmed numerically that the logarithmic coefficient at $d=5$ is consistent with the analytical prediction for $A_5$.

This result shows that the presence of long-range correlations does not necessarily imply a macroscopically large mutual information. In particular, for $d<5$, the mutual information shared between the left and right regions remains $O(1)$. This dimension dependence originates from the $\mathcal{I}_{\mathrm{hyd}, \boldsymbol\alpha}^{(d)}(n) \sim |\bm m|^{-4}$ decay of the contribution from each transverse mode together with the density of transverse modes.

Furthermore, writing $n=L/h$ makes the dependence of the mutual information on the microscopic cutoff explicit. For $d<5$, the leading behavior is independent of $h$; at $d=5$, a logarithmic dependence $\log(L/h)$ appears; and for $d>5$, the leading behavior itself becomes cutoff dependent. Thus, $d=5$ is not only the critical dimension at which the scaling of the mutual information changes, but also the boundary between a regime in which the leading behavior is fully determined by long-wavelength hydrodynamics and one in which microscopic information is required.

In a heat conductor described by a short-range Hamiltonian, microscopic
equilibrium correlations across the cut can give rise to mutual information
proportional to the interfacial area. If local equilibrium holds near
the cut, we expect this area term to persist in the conducting steady state and
to dominate the hydrodynamic contribution found here for $d\geq 2$.

Our analysis has been restricted to linear Gaussian fluctuating hydrodynamics with a single conserved energy field. It remains an open problem to determine how the scaling of mutual information and the associated critical dimension are modified in systems with multiple conserved quantities, nonlinear fluctuating hydrodynamics, or anomalous transport.

\begin{acknowledgments}
RK was supported by JST SPRING Grant Number JPMJSP2110. This work was supported by JSPS KAKENHI Grant Nos.~JP26H02015 and JP26H00388.

\end{acknowledgments}

\section{Stationary Independence between Different Transverse Modes}
\label{app:independence}
Here, we show that the stationary cross-covariance between different real transverse modes vanishes, implying that different transverse modes are independent of each other in the steady state.

For two distinct transverse modes $\boldsymbol\alpha$ and $\boldsymbol\beta$, we define the stationary cross-covariance as
\begin{align}
C_{\boldsymbol\alpha\boldsymbol\beta}
:=
\left\langle
\delta\hat{\widetilde{\bm u}}_{\boldsymbol\alpha}
\delta\hat{\widetilde{\bm u}}_{\boldsymbol\beta}^{\mathsf T}
\right\rangle_{\mathrm{ns}}
\end{align}

From Eq.~\eqref{eq:modeDynamics}, $C_{\boldsymbol\alpha\boldsymbol\beta}$ satisfies
\begin{align}
0={}&
-D\Delta_{\boldsymbol\alpha}
C_{\boldsymbol\alpha\boldsymbol\beta}
-D
C_{\boldsymbol\alpha\boldsymbol\beta}
\Delta_{\boldsymbol\beta}
+
2\kappa
\left\langle
\hat{\bm\eta}_{\boldsymbol\alpha}(t)
\hat{\bm\eta}_{\boldsymbol\beta}^{\mathsf T}(t)
\right\rangle
\end{align}
where the last term originates from the equal-time correlation of the white noise.

From Eq.~\eqref{eq:modeNoiseCovariance}, for $\boldsymbol\alpha\neq\boldsymbol\beta$, we have
\begin{align}
\left\langle
\hat{\bm\eta}_{\boldsymbol\alpha}(t)
\hat{\bm\eta}_{\boldsymbol\beta}^{\mathsf T}(t')
\right\rangle
=0
\end{align}
Therefore, in the steady state, the following relation holds:
\begin{align}
\Delta_{\boldsymbol\alpha}
C_{\boldsymbol\alpha\boldsymbol\beta}
+
C_{\boldsymbol\alpha\boldsymbol\beta}
\Delta_{\boldsymbol\beta}
=0
\label{eq:crossModeSylvester}
\end{align}

Here, we use the fact that $\Delta_{\boldsymbol\alpha}>0$ and $\Delta_{\boldsymbol\beta}>0$. Multiplying Eq.~\eqref{eq:crossModeSylvester} from the left by $C_{\boldsymbol\alpha\boldsymbol\beta}^{\mathsf T}$ and taking the trace yields
\begin{align}
0
={}&
\mathrm{tr}
\left[
C_{\boldsymbol\alpha\boldsymbol\beta}^{\mathsf T}
\Delta_{\boldsymbol\alpha}
C_{\boldsymbol\alpha\boldsymbol\beta}
\right]
\nonumber\\
&+
\mathrm{tr}
\left[
C_{\boldsymbol\alpha\boldsymbol\beta}^{\mathsf T}
C_{\boldsymbol\alpha\boldsymbol\beta}
\Delta_{\boldsymbol\beta}
\right]
\nonumber\\
={}&
\left\|
\Delta_{\boldsymbol\alpha}^{1/2}
C_{\boldsymbol\alpha\boldsymbol\beta}
\right\|_{\mathrm F}^{2}
+
\left\|
C_{\boldsymbol\alpha\boldsymbol\beta}
\Delta_{\boldsymbol\beta}^{1/2}
\right\|_{\mathrm F}^{2},
\end{align}
where $\Vert{}\cdot\Vert{}{\mathrm F}$ denotes the Frobenius norm. Since each term on the right-hand side is non-negative, we must have
\begin{align}
C_{\boldsymbol\alpha\boldsymbol\beta}=0,
\qquad
\boldsymbol\alpha\neq\boldsymbol\beta
\end{align}

\section{Details of the Uniform Bounds on the Single-Mode Mutual Information}
\label{app:modeMIBound}

We provide the details of the uniform bound on the single-mode mutual information used in Sec.~VI.

\subsection{Bounds on the Self-Covariance Blocks}

Restating Eq.~\eqref{eq:canonicalCorrelation} derived in Sec.~III, the mutual information of each transverse mode $\balpha$ is given by
\begin{align}
\mathcal{I}_{\mathrm{hyd},\balpha}^{(d)}(n)
&=
-\frac{1}{2}
\log\left(1-\sigma_{\balpha}^2\right),
\nonumber\\
\sigma_{\balpha}^2
&=
\frac{
\bms_{\balpha}^{\mathsf T}
A_{\balpha}^{-1}
\bms_{\balpha}
}{
s_{\balpha,2n+1}
}
\frac{
\bm t_{\balpha}^{\mathsf T}
B_{\balpha}^{-1}
\bm t_{\balpha}
}{
s_{\balpha,2n+1}
}
\end{align}

Here, $\bms_{\balpha}$ and $\bm t_{\balpha}$ are the vectors defined in Eq.~\eqref{eq:sVector}, and $A_{\balpha}$ and $B_{\balpha}$ are the autocovariance blocks corresponding to the left and right regions of the covariance $\Sigma_{\balpha}$. That is, from Eqs.~\eqref{eq:GammaAlpha} and \eqref{eq:LRblock}, they can be written as
\begin{align}
\Sigma_{\balpha}
&=\frac{\kappa\delta_n^2}{D}
\left(Q+G_{\balpha} \right),
\qquad
G_{\balpha}:=\Delta_{\balpha}^{-1},
\\
\Sigma_{\balpha}
&=\frac{\kappa\delta_n^2}{D}
\begin{pmatrix}
A_{\balpha} & C_{\balpha}\\
C_{\balpha}^{\mathsf T} & B_{\balpha}
\end{pmatrix}.
\end{align}

Here, $Q$ is the diagonal matrix defined in Eq.~\eqref{eq:Qdef}, and partitioning $G_{\balpha}$ into blocks corresponding to the left and right regions yields
\begin{align}
A_{\balpha}
&=
Q_L+G_{\balpha}^{LL},
\nonumber\\
B_{\balpha}
&=
Q_R+G_{\balpha}^{RR}
\label{eq:normalizedSelfCovBlocks}
\end{align}
where $Q_L$ and $Q_R$ are the principal submatrices of $Q$ corresponding to the left and right regions, respectively. In the following, we clarify the wavenumber dependence of $\sigma_{\balpha}$ by evaluating these autocovariance blocks uniformly.

To estimate $A_{\balpha}$ and $B_{\balpha}$, we first consider
the operator norm of $G_{\balpha}$.
Since $\rho_{\balpha}\ge0$, we have
\begin{align}
\Delta_{\balpha}
=
L_N^{D}
+
\rho_{\balpha}\Id_N
\ge
L_N^{D}.
\end{align}
The smallest eigenvalue of the Dirichlet Laplacian with $N=2n$ is
\begin{align}
\lambda_{\min}(L_N^{D})
=
4\sin^2
\left(
\frac{\pi}{2(2n+1)}
\right)
\ge
\frac{4}{(2n+1)^2}.
\end{align}
Therefore,
\begin{align}
\left\|
G_{\balpha}
\right\|
=
\left\|
\Delta_{\balpha}^{-1}
\right\|
\le
\frac{(2n+1)^2}{4}.
\end{align}
Since $G_{\balpha}$ is positive definite, its diagonal blocks satisfy
\begin{align}
0
\le
G_{\balpha}^{LL},
G_{\balpha}^{RR}
\le
\|G_{\balpha}\|\Id_n.
\end{align}
On the other hand, $T_\ell\le T_z\le T_r$ implies
\begin{align}
\frac{T_\ell^2}{\delta_n^2}\Id_n
\le
Q_L,Q_R
\le
\frac{T_r^2}{\delta_n^2}\Id_n.
\end{align}
Combining these bounds, we obtain
\begin{align}
\begin{split}
\frac{T_\ell^2}{\delta_n^2}\Id_n
\le
A_{\balpha},B_{\balpha}
\le
\frac{M_T}{\delta_n^2}\Id_n,
\\
M_T
:=
T_r^2+\frac{(T_r-T_\ell)^2}{4}
\end{split}
\label{eq:selfCovUniformBound}
\end{align}

Therefore,
\begin{align}
\frac{
\delta_n^2
\|\bm s_{\balpha}\|^2
}{
M_T
}
&\le
\bm s_{\balpha}^{\mathsf T}
A_{\balpha}^{-1}
\bm s_{\balpha}
\le
\frac{
\delta_n^2
\|\bm s_{\balpha}\|^2
}{
T_\ell^2
},
\nonumber\\
\frac{
\delta_n^2
\|\bm t_{\balpha}\|^2
}{
M_T
}
&\le
\bm t_{\balpha}^{\mathsf T}
B_{\balpha}^{-1}
\bm t_{\balpha}
\le
\frac{
\delta_n^2
\|\bm t_{\balpha}\|^2
}{
T_\ell^2
}.
\end{align}
Since $\|\bm t_{\balpha}\|=\|\bm s_{\balpha}\|$
by Eq.~\eqref{eq:sVector},
the wave-number dependence of $\sigma_{\balpha}$
is reduced to estimating
$\|\bm s_{\balpha}\|^2/s_{\balpha,2n+1}$.
We therefore define
\begin{align}
R_n(\gamma)
:=
\frac{
\displaystyle
\sum_{k=1}^{n}
s_k(\gamma)^2
}{
s_{2n+1}(\gamma)
}
\label{eq:RnDef}
\end{align}
Then, we obtain
\begin{align}
\frac{\delta_n^2}{M_T}
R_n(\gamma_{\balpha})
\le
\sigma_{\balpha}
\le
\frac{\delta_n^2}{T_\ell^2}
R_n(\gamma_{\balpha})
\label{eq:sigmaRnBound}
\end{align}

\subsection{Two-Sided Bounds on the Hyperbolic Sum and Transverse Wave Numbers}

The quantity $R_n(\gamma)$ can be evaluated analytically.
For $\gamma>0$, we have
\begin{align}
\sum_{k=1}^{n}\cosh(2k\gamma)
=
\frac{
\sinh((2n+1)\gamma)-\sinh\gamma
}{
2\sinh\gamma
}.
\end{align}
Hence,
\begin{align}
\sum_{k=1}^{n}s_k(\gamma)^2
=
\frac{
\sinh((2n+1)\gamma)-(2n+1)\sinh\gamma
}{
4\sinh^3\gamma
}.
\end{align}
Therefore, Eq.~\eqref{eq:RnDef} gives
\begin{align}
R_n(\gamma)
=
\frac{1}{4\sinh^2\gamma}
\left[
1-
\frac{
(2n+1)\sinh\gamma
}{
\sinh((2n+1)\gamma)
}
\right].
\label{eq:RnExact}
\end{align}
We introduce
\begin{align}
B_n(\gamma)
:=
1-
\frac{
(2n+1)\sinh\gamma
}{
\sinh((2n+1)\gamma)
}
\label{eq:BnDef}
\end{align}
Since $0<B_n(\gamma)<1$, an upper bound on $R_n$
follows immediately.
In the following, we establish a positive
lower bound on $B_n(\gamma_{\balpha})$
for all nonzero transverse modes.

For any nonzero transverse mode, at least one
$m_a$ is nonzero, and hence
\begin{align}
\rho_{\balpha}
\ge
4\sin^2\left(\frac{\pi}{n}\right).
\end{align}
On the other hand,
$\cosh\gamma_{\balpha}=1+\rho_{\balpha}/2$ implies
\begin{align}
\sinh^2
\left(
\frac{\gamma_{\balpha}}{2}
\right)
=
\frac{\rho_{\balpha}}{4}
\ge
\sin^2\left(\frac{\pi}{n}\right).
\end{align}
Using
$\operatorname{arsinh}x\ge x/\sqrt{2}$
for $0\le x\le1$ and
$\sin(\pi/n)\ge2/n$ for $n\ge2$, we obtain
\begin{align}
\gamma_{\balpha}
\ge
\sqrt{2}\sin\left(\frac{\pi}{n}\right)
\ge
\frac{2\sqrt{2}}{n}.
\end{align}
Therefore, for all nonzero transverse modes,
\begin{align}
(2n+1)\gamma_{\balpha}>4
\label{eq:gammaLengthLower}
\end{align}
holds.

We now define
\begin{align}
H(x):=\frac{\sinh x}{x}.
\end{align}
Then,
\begin{align}
\frac{
(2n+1)\sinh\gamma
}{
\sinh((2n+1)\gamma)
}
=
\frac{
H(\gamma)
}{
H((2n+1)\gamma)
}.
\end{align}
Let $x_{\balpha}:=(2n+1)\gamma_{\balpha}$.
For $n\ge2$, we have
$\gamma_{\balpha}\le x_{\balpha}/5$.
Since $H$ is monotonically increasing and
$\log H$ is convex, the ratio
$H(x/5)/H(x)$ is monotonically decreasing
for $x>0$.
Thus, using Eq.~\eqref{eq:gammaLengthLower},
we obtain
\begin{align}
\frac{
H(\gamma_{\balpha})
}{
H((2n+1)\gamma_{\balpha})
}
\le
\frac{
H(x_{\balpha}/5)
}{
H(x_{\balpha})
}
\le
\frac{H(4/5)}{H(4)}.
\end{align}
Defining
\begin{align}
b_0
:=
1-\frac{H(4/5)}{H(4)}
>0,
\end{align}
we find
\begin{align}
b_0
\le
B_n(\gamma_{\balpha})
<1.
\end{align}
Consequently, for all nonzero transverse modes,
we obtain the uniform two-sided bound
\begin{align}
\frac{
b_0
}{
4\sinh^2\gamma_{\balpha}
}
\le
R_n(\gamma_{\balpha})
\le
\frac{
1
}{
4\sinh^2\gamma_{\balpha}
}
\label{eq:RnTwoSided}
\end{align}

Next, we estimate the transverse-wave-number dependence of
$\sinh^2\gamma_{\balpha}$.
Let $\bm m=(m_2,\ldots,m_d)$ denote the integer wave-vector
index corresponding to the real transverse mode
$\balpha=(\alpha_2,\ldots,\alpha_d)$,
and define $r:=|\bm m|$.
For each $a=2,\ldots,d$, we have
\begin{align}
0
\le
\frac{\pi m_a}{n}
\le
\frac{\pi}{2}.
\end{align}
Since $2x/\pi\le\sin x\le x$ holds in this interval,
we obtain
\begin{align}
\frac{2m_a}{n}
\le
\sin\left(\frac{\pi m_a}{n}\right)
\le
\frac{\pi m_a}{n}.
\end{align}
Therefore, from
\begin{align}
\rho_{\balpha}
=
4\sum_{a=2}^{d}
\sin^2\left(\frac{\pi m_a}{n}\right),
\end{align}
we obtain
\begin{align}
\frac{16r^2}{n^2}
\le
\rho_{\balpha}
\le
4\pi^2\frac{r^2}{n^2}.
\label{eq:rhoRBound}
\end{align}
Furthermore,
$\cosh\gamma_{\balpha}=1+\rho_{\balpha}/2$ implies
\begin{align}
\sinh^2\gamma_{\balpha}
=
\rho_{\balpha}
\left(
1+\frac{\rho_{\balpha}}{4}
\right).
\end{align}
Since $\rho_{\balpha}\le4(d-1)$,
combining this with Eq.~\eqref{eq:rhoRBound} yields
\begin{align}
\frac{16r^2}{n^2}
\le
\sinh^2\gamma_{\balpha}
\le
4\pi^2d\frac{r^2}{n^2}.
\label{eq:gammaRBound}
\end{align}

Finally, we substitute Eqs.~\eqref{eq:RnTwoSided}
and \eqref{eq:gammaRBound} into
Eq.~\eqref{eq:sigmaRnBound}.
For the upper bound, we obtain
\begin{align}
\sigma_{\balpha}
\le
\frac{(T_r-T_\ell)^2}{256T_\ell^2}
\frac{1}{r^2}.
\end{align}
For the lower bound, we obtain
\begin{align}
\sigma_{\balpha}
\ge
\frac{
b_0(T_r-T_\ell)^2
}{
144\pi^2dM_T
}
\frac{1}{r^2}.
\end{align}
Here, we have used
$\delta_n=(T_r-T_\ell)/(2n+1)$ and
\begin{align}
\frac{1}{3}
\le
\frac{n}{2n+1}
<
\frac{1}{2}.
\end{align}
Thus, defining
\begin{align}
c_\sigma
&:=
\frac{
b_0(T_r-T_\ell)^2
}{
144\pi^2dM_T
},
\nonumber\\
C_\sigma
&:=
\frac{
(T_r-T_\ell)^2
}{
256T_\ell^2
},
\end{align}
we obtain the uniform two-sided bound
\begin{align}
\frac{c_\sigma}{r^2}
\le
\sigma_{\balpha}
\le
\frac{C_\sigma}{r^2}
\label{eq:sigmaFinalTwoSided}
\end{align}
for all nonzero transverse modes.

\subsection{Single-Mode Mutual Information and the Zero Mode}

To translate the bound in Eq.~\eqref{eq:sigmaFinalTwoSided}
into a bound on the mutual information, we first show that
$\sigma_{\balpha}$ is uniformly bounded away from $1$
for all transverse modes.
Since $A_{\balpha}\ge Q_L$ and $B_{\balpha}\ge Q_R$
by Eq.~\eqref{eq:normalizedSelfCovBlocks}, we have
\begin{align}
A_{\balpha}^{-1}\le Q_L^{-1},
\qquad
B_{\balpha}^{-1}\le Q_R^{-1}.
\end{align}
Furthermore,
\begin{align}
q_z
=
\left(\frac{T_z}{\delta_n}\right)^2
=
\left(\frac{T_\ell}{\delta_n}+z\right)^2
\ge z^2.
\end{align}
Since $H(x)=\sinh x/x$ is monotonically increasing,
we have
\begin{align}
\frac{s_{\balpha,i}}{s_{\balpha,n}}
\le
\frac{i}{n}
\end{align}
for $1\le i\le n$.
Therefore,
\begin{align}
\bm s_{\balpha}^{\mathsf T}
A_{\balpha}^{-1}
\bm s_{\balpha}
&\le
\bm s_{\balpha}^{\mathsf T}
Q_L^{-1}
\bm s_{\balpha}
\nonumber\\
&\le
\sum_{i=1}^{n}
\frac{s_{\balpha,i}^2}{i^2}
\le
\frac{s_{\balpha,n}^2}{n}.
\end{align}
Similarly, using
$t_{\balpha,i}\le s_{\balpha,n}$ and
$q_{n+i}\ge(n+i)^2\ge n^2$, we obtain
\begin{align}
\bm t_{\balpha}^{\mathsf T}
B_{\balpha}^{-1}
\bm t_{\balpha}
\le
\frac{s_{\balpha,n}^2}{n}.
\end{align}
Thus, Eq.~\eqref{eq:canonicalCorrelation} gives
\begin{align}
\sigma_{\balpha}
\le
\frac{s_{\balpha,n}^2}
{n\,s_{\balpha,2n+1}}.
\end{align}
Furthermore,
\begin{align}
\frac{s_{\balpha,n}^2}
{n\,s_{\balpha,2n+1}}
=
\frac{n}{2n+1}
\frac{
H(n\gamma_{\balpha})^2
}{
H(\gamma_{\balpha})
H((2n+1)\gamma_{\balpha})
}.
\end{align}
The monotonicity of $H$ and the convexity of
$\log H$ imply
\begin{align}
H(n\gamma_{\balpha})^2
\le
H(\gamma_{\balpha})
H((2n+1)\gamma_{\balpha}).
\end{align}
Consequently,
\begin{align}
0
\le
\sigma_{\balpha}
\le
\frac{n}{2n+1}
<
\frac12.
\label{eq:sigmaUniformHalf}
\end{align}

We now apply the elementary inequality
\begin{align}
x
\le
-\log(1-x)
\le
\frac{x}{1-x},
\end{align}
valid for $0\le x<1$, with
$x=\sigma_{\balpha}^2$.
Since Eq.~\eqref{eq:sigmaUniformHalf} implies
$\sigma_{\balpha}^2\le1/4$, we obtain
\begin{align}
\frac12\sigma_{\balpha}^2
\le
\mathcal{I}_{\mathrm{hyd}, \boldsymbol\alpha}^{(d)}(n)
\le
\frac23\sigma_{\balpha}^2.
\label{eq:MIvsSigma}
\end{align}
Therefore, for all nonzero transverse modes,
\begin{align}
\frac{c_\sigma^2}{2r^4}
\le
\mathcal{I}_{\mathrm{hyd}, \boldsymbol\alpha}^{(d)}(n)
\le
\frac{2C_\sigma^2}{3r^4}.
\end{align}

Finally, we consider the zero transverse mode
$\balpha=\bm0$.
For this mode, $\rho_{\bm0}=0$ implies
$\gamma_{\bm0}=0$, and hence
$s_{\bm0,k}=k$.
Therefore,
\begin{align}
R_n(0)
=
\frac{\sum_{k=1}^{n}k^2}{2n+1}
=
\frac{n(n+1)}{6}.
\label{eq:RnZero}
\end{align}
Since Eq.~\eqref{eq:sigmaRnBound} also applies
to the zero mode, defining
$\Delta T:=T_r-T_\ell$, we obtain
\begin{align}
\frac{\Delta T^2}{6M_T}
\frac{n(n+1)}{(2n+1)^2}
\le
\sigma_{\bm0}
\le
\frac{\Delta T^2}{6T_\ell^2}
\frac{n(n+1)}{(2n+1)^2}.
\end{align}
Since
$n(n+1)/(2n+1)^2\to1/4$ as $n\to\infty$,
we find, for fixed $\Delta T>0$,
\begin{align}
\sigma_{\bm0}
=
\Theta(1).
\end{align}
Furthermore, Eq.~\eqref{eq:MIvsSigma} gives
\begin{align}
I_{\bm0}(L:R)
=
\Theta(1).
\label{eq:zeroModeMI}
\end{align}
Thus, the zero transverse mode contributes
a finite, nonzero amount of mutual information
independent of the system size, whereas
$\mathcal{I}_{\mathrm{hyd}, \boldsymbol\alpha}^{(d)}(n)
\asymp r^{-4}$
holds uniformly for all nonzero transverse modes.
This establishes Eq.~\eqref{eq:mainModeMI}.
Finally, applying the standard counting argument
for transverse modes to Eq.~\eqref{eq:MIsum}
yields Eq.~\eqref{eq:mainTotalMI}.

\twocolumngrid
\bibliography{refs_grouped_by_topic}

\clearpage
\appendix

\end{document}